\documentclass[twocolumn]{aastex62}
\usepackage{xspace,graphics,graphicx,booktabs}
\newcommand\msun{\ensuremath{M_\sun}\xspace}
\newcommand\teff{\ensuremath{T_{\rm eff}}\xspace}
\newcommand\logg{\ensuremath{\log g}\xspace}
\submitjournal{The Astrophysical Journal}
\accepted{2018 July 24}
\setwatermarkfontsize{1in}
\shorttitle{White Dwarfs in Messier 67}
\shortauthors{Williams et al.}

\begin{document}
\title{Ensemble Properties of the White Dwarf Population of the Old, Solar Metallicity Open Star Cluster Messier 67\footnote{Some of the data presented herein were obtained at the W.M. Keck Observatory, which is operated as a scientific partnership among the California Institute of Technology, the University of California and the National Aeronautics and Space Administration. The Observatory was made possible by the generous financial support of the W.M. Keck Foundation.}}

\author[0000-0002-1413-7679]{Kurtis A. Williams}
\affiliation{Department of Physics \& Astronomy, Texas A\&M University-Commerce, P.O. Box 3011, Commerce, TX, 75429-3011, USA}

\author{Paul A. Canton}
\affiliation{Homer L.~Dodge Department of Physics and Astronomy, University of Oklahoma, 440 W.~Brooks St., Normon, OK 73019, USA}

\author[0000-0003-3858-637X]{A.\ Bellini}
\affiliation{Space Telescope Science Institute, 3700 San Martin Dr., Baltimore, MD 21218, USA}

\author[0000-0001-5296-6251]{Michael Bolte}
\affiliation{UCO/Lick Observatory, University of California, 1156 High St., Santa Cruz, CA 95064, USA}

\author{Kate H. R. Rubin}
\affiliation{Department of Astronomy, San Diego State University, San Diego, CA 92182, USA}

\author[0000-0002-8655-4308]{Alexandros Gianninas}
\affiliation{Homer L.~Dodge Department of Physics and Astronomy, University of Oklahoma, 440 W.~Brooks St., Normon, OK 73019, USA}

\author[0000-0001-6098-2235]{Mukremin Kilic}
\affiliation{Homer L.~Dodge Department of Physics and Astronomy, University of Oklahoma, 440 W.~Brooks St., Normon, OK 73019, USA}

\correspondingauthor{Kurtis A. Williams}
\email{Kurtis.Williams@tamuc.edu}

\begin{abstract}
White dwarfs are excellent forensic tools for studying end-of-life issues surrounding low- and intermediate-mass stars, and the old, solar-metallicity open star cluster Messier 67 is a proven laboratory for the study of stellar evolution for solar-type stars.  In this paper, we present a detailed spectroscopic study of brighter ($M_g\leq 12.4$) white dwarfs in Messier 67, and, in combination with previously-published proper motion membership determinations, we identify a clean, representative sample of cluster white dwarfs, including 13 members with hydrogen-dominated atmospheres, at least one of which is a candidate double degenerate, and 5 members with helium-dominated atmospheres.  Using this sample we test multiple predictions surrounding the final stages of stellar evolution in solar type stars.  In particular, the stochasticity of the integrated mass lost by $\sim 1.5$ solar mass stars is less than 7\% of the white dwarf remnant mass. We identify white dwarfs likely resulting from binary evolution, including at least one blue straggler remnant and two helium core white dwarfs.  We observe no evidence of a significant population of helium core white dwarfs formed by enhanced mass loss on the red giant branch of the cluster.  The distribution of white dwarf atmospheric compositions is fully consistent with that in the field, limiting proposed mechanisms for the suppression of helium atmosphere white dwarf formation in star clusters.  In short, the white dwarf population of Messier 67 is fully consistent with basic predictions of single- and multiple-star stellar evolution theories for solar metallicity stars.
\end{abstract}


\defcitealias{PCanton2017}{Paper~II}

\section{Introduction \label{sec.intro}}

Intermediate-age and old open star clusters (ages $\gtrsim 50$ Myr) are ideal laboratories for studying the late stages of intermediate- and low-mass stellar evolution.  The stars within a given open cluster are coeval and nearly identical in elemental abundances, with most properties dependent primarily on the stellar mass.  White dwarfs (WDs) are the remnants of stars with masses $M\lesssim 8 M_\sun$.  Therefore, the study of WD populations in open star clusters can be an exceptionally effective means of studying the late stages of stellar evolution, as the progenitor star properties of age, metallicity, and mass can be constrained tightly.   In this paper, we present the results of combined photometric, spectroscopic and astrometric studies of the WD population in the old open star cluster \object{Messier 67} (M67, NGC 2682).  The results of this study are useful in probing several different areas of stellar and WD evolution.

\subsection{White Dwarf Masses \label{sec.intro.mass}}
Much effort in the study of open cluster WD populations has gone into understanding the relationship between WD mass and the derived progenitor star mass, known as the initial-final mass relation (IFMR). A proper review of the IFMR literature and the contribution of the M67 WD population to the semi-empirical IFMR are given in a companion manuscript by \citeauthor{PCanton2017} \citepalias[2017; hereafter][]{PCanton2017}.  

A topic closely related to the IFMR is the intrinsic scatter in the relationship.  Given a fixed progenitor star mass and metallicity, is there any stochasticity in the resulting WD mass?  Basic single star evolutionary theory predicts stellar evolution is uniquely determined by the zero age main sequence mass and metallicity of a star, yet it is plausible that some randomness in the remnant WD mass could be  introduced by events such as core mass loss during dredge up, the number of thermal pulses during the asympotic giant branch phase, and the rate and timing of post main sequence mass loss.  For such studies, a population of WDs arising from stars of nearly identical masses would be highly useful, especially for solar-metallicity stars \citep[some spectroscopic globular cluster WD mass measurements exist, e.g.,][]{2004A&A...420..515M,2009ApJ...705..408K}.  

M67 is among the oldest known open star clusters.  Its age is typically quoted as being $\approx 4$ Gyr \citep[e.g.,][]{1993AJ....106..181M,1998ApJ...504L..91R,2004PASP..116..997V,2009ApJ...698.1872S,2010A&A...513A..50B}, though the most recent studies using a variety of age dating techniques such as asteroseismology \citep{2016ApJ...832..133S}, main sequence fitting \citep{2015MNRAS.450.2500B}, and modified stellar models \citep{2014MNRAS.444.2525C} lean toward a younger age of $\approx 3.5$ Gyr.  

For this paper, the exact age is not crucial --  the zero age main sequence mass of a star at the tip of the AGB (i.e., those stars currently forming WDs) is only slowly evolving in time, changing by $\lesssim 0.05 M_\sun$ \citep{2015MNRAS.452.1068C} between cluster ages of 3.5 Gyr and 4.0 Gyr.  More important in our study is the change in progenitor mass for WDs in our sample.  The WDs used in the analysis in this paper have cooling ages $\lesssim 700$ Myr, and therefore have a spread in zero-age main sequence mass of $\lesssim 0.1 M_\sun$.  Cluster WDs from more massive progenitors exist, but are fainter than the photometric limits of this study.  We therefore will be able to measure the intrinsic spread in WD mass for stars of virtually identical composition and initial mass.

Another ancient metal-rich star cluster with a rich, spectroscopically well-studied WD sequence is \object{NGC 6791}, which has a super-solar metallicity ($\left[ \mathrm{Fe/H} \right] \approx+0.4$, e.g., \citealt{2006ApJ...642..462G,2006ApJ...646..499O}) and an age of $\sim 8$ Gyr.   The spectroscopic WD study of \citet{Kalirai2007} found a majority of cluster WDs with masses $\lesssim 0.5 M_\sun$, consistent with these WDs harboring helium cores instead of the canonical carbon-oxygen cores formed by helium burning during the asymptotic giant branch phases of stellar evolution.  Although He core WDs can be formed by binary star evolution (see Section \ref{sec.intro.binary}), the binary fraction of NGC 6791 is too low to explain the large fraction of He core WDs in this cluster.  

\citet{2005ApJ...635..522H} and \citet{Kalirai2007} hypothesize that the high metallicity of NGC 6791 may lead to excessive mass loss from stars ascending the red giant branch, resulting in stellar cores too low in  mass to undergo a helium flash. \citet{2007ApJ...671..761K}  similarly invoke this mechanism to explain the large number of low-mass single WDs in the field.  Enhanced mass loss on the red giant branch is also invoked to explain the large number of extreme horizontal branch stars in NGC 6791 noted by \citet{1994AJ....107.1408L} as red giants that retained just enough mass for a helium flash, and to explain the bimodal white dwarf luminosity function (WDLF) reported by \citet{2005ApJ...624L..45B,2008ApJ...678.1279B}.  Other mechanisms such as double degenerates  have been proposed to explain NGC 6791's peculiar WDLF \citep{2008ApJ...679L..29B,2010Natur.465..194G}, and  little evidence of ongoing enhanced red giant mass loss has been observed \citep{2008ApJ...680L..49V,2012MNRAS.419.2077M}. 

With its solar metallicity, lack of extreme horizontal branch stars \citep{1994AJ....107.1408L}, and unimodal WDLF \citep{1998ApJ...504L..91R,2010A&A...513A..50B}, the enhanced mass loss mechanism would suggest that M67 should have a lower He-core WD fraction than NGC 6791; a detailed study of WD masses to determine this fraction is therefore in order, especially since recent asteroseismological studies of M67 red giants has found no evidence of strong mass loss \citep{2016ApJ...832..133S}.

\subsection{Products of Binary Star Evolution \label{sec.intro.binary}}

Open star clusters have significant binary star fractions; M67 itself has a main sequence binary fraction of $\gtrsim 38\%$ \citep[e.g.][]{1993AJ....106..181M,1996AJ....112..628F}.  Signs of binary interaction are common among open cluster member populations.  Blue stragglers are stars with masses significantly higher than the main sequence turnoff mass; these are thought to be the result of significant mass transfer \citep[e.g.,][]{2008MNRAS.387.1416C,2011Natur.478..356G,2014ApJ...783L...8G} and binary coalescence \citep[e.g.,][]{2002ApJ...570..184H,2005MNRAS.363..293H,2009ApJ...697.1048P}.  If the mass transfer / merger results in significant mixing of unburned hydrogen into the stellar core, in essence resetting the clock on stellar evolution, a blue straggler with a mass greater than the main sequence turnoff mass will produce a more massive WD than the single stars completing their evolution at that point in time.

M67 is known to have a significant blue straggler population \citep[e.g.,][]{1986AJ.....92.1364M,1992PASP..104.1268M,1998AJ....116..789L}, including evolved blue stragglers such as \object[Cl* NGC 2682 SAND 1237]{S1237} \citep{2016ApJ...832L..13L} and \object[Cl* NGC 2682 SAND 1040]{S1040} \citep{1997ApJ...481L..93L}.  Given the presence of evolved blue stragglers in the cluster, it is highly likely that additional blue straggler stars have completed their evolution and evolved into WD remnants.  Further, since many blue stragglers have low-mass, helium-core WD companions, the WD progeny of blue straggler stars are likely to appear to be single WDs.  This is because low-mass WDs have a large surface area and He-core WDs have significant lower heat capacity than the typical carbon-oxygen core WDs, resulting in He-core WDs cooling and fading below our detection limits quite rapidly.

The presence of helium-core WDs in an open cluster can be another sign of binary interactions.  The minimum core mass required for helium ignition in an evolved star is $\approx 0.45 M_\sun$ \citep[e.g.,][]{1994ApJ...426..612S,2001PASP..113..409F}.  WDs more massive than this limit are composed primarily of carbon and oxygen, the primary results of core helium burning.  Less massive WDs, not having ignited helium, should have core compositions of helium.  However, stellar evolution models predict that the nuclear lifetime of a single star with insufficient mass for core helium ignition is significantly longer than a Hubble time, therefore any He-core WDs in the present day universe cannot have arisen by standard stellar evolution.  Close binary evolution is one source of He-core WDs; if a common envelope forms around a red giant star and a companion, the envelope can be ejected and nuclear evolution halted prior to the helium flash\citep[e.g.,][]{1993PASP..105.1373I}.  The fact that a large fraction of low-mass He-core WDs in the field are short-period binary systems is strong evidence of this mechanism \citep[e.g.,][]{1995MNRAS.275..828M,2010ApJ...723.1072B,2015ApJ...812..167G}.

Finally, we note that one cataclysmic variable system is known in M67, \object{EU Cnc} \citep{1991AJ....101..541G,1994AA...290L..17P}.  Our observations and analysis of this system are presented in \citet{2013AJ....145..129W} and are not discussed further in this paper.

\subsection{White Dwarf Atmospheric Composition \label{sec.intro.atm}}

The high surface gravity of WDs results in highly stratified atmospheres usually dominated by a single atomic species.  The two most common atmospheric types of WDs are the hydrogen-dominated atmospheres (spectral type DA) and helium-dominated atmospheres (almost all non-DA spectral types).  The accepted formation scenario of most non-DA WDs is the ``born again'' scenario, in which a late thermal pulse results in the loss of the WD's hydrogen layer \citep{1983ApJ...264..605I}.   

The ratio of DA to non-DA WDs is a function of temperature, with most changes explicable by the mixing or gravitational separation of atmospheric layers by the competing effects of diffusion and changing convection zone depth \citep[e.g.,][]{1987fbs..conf..319F,2011ApJ...737...28B}, but for warmer WDs such as those  presented in this paper, the fraction of DA WDs is $\approx 80\%$ \citep[e.g.,][]{2015A&A...583A..86K}.

\citet{2005ApJ...618L.129K} and \citet{2009ApJ...705..398D} have proposed that some intermediate-age open star clusters and some globular clusters exhibit a dearth of non-DA WDs, requiring a mechanism suppressing non-DA formation in the star cluster environment.  However, some non-DA WDs are known to exist in intermediate-age open star clusters at a fraction comparable to the field if the ``DB gap" and the different cooling rates of DA and non-DA WDs are appropriately considered \citep{2006ApJ...643L.127W}.  Again, a large sample of WDs with solar-metallicity progenitors, like we present in this paper, could be key to proper determination of whether open cluster environments exhibit any difference of DA to non-DA ratio from the field population.  

\subsection{Improved Cluster Membership Determination\label{sec.intro.pms}}

In most work involving open cluster WDs, cluster membership is determined solely by cuts in apparent distance modulus \citep[e.g.,][]{2009ApJ...693..355W,Dobbie2012,2016ApJ...818...84C}.  \citet{Dobbie2009}, \citet{2012A&A...547A..99T}, and \citet{2010A&A...513A..50B}, among others, demonstrate that proper motion membership determinations are crucial in investigating open cluster WD populations.  Proper motion memberships not only reliably reduce field contamination in IFMR studies, but they also permit identification of other astrophysically interesting WDs for which knowledge of progenitor mass, system age, and primordial metallicity are important but typically unknown quantities.  These include WDs with differing atmospheric compositions, cataclysmic variables, and double degenerates \citep[e.g.,][]{2006ApJ...643L.127W,2013AJ....145..129W,2015AJ....150..194W}.  However, most open clusters lack precise proper motions for faint stars such as WDs; most open cluster WDs located at distances $\gtrsim 1$ kpc are fainter than the likely ultimate limits of GAIA astrometry ($V\lesssim 21.5$).  

Precision proper motion membership measurements for open cluster WDs  requires multi-epoch wide field imaging of clusters using $4$ m class telescopes or larger.  After the advent of large-format CCDs in the mid-to-late 1990s, a first epoch of deep open cluster imaging dedicated to WD studies was obtained by multiple groups \citep[e.g.][]{2001ApJ...563..987C,2001AJ....122..257K,Williams2002}.  Since then, enough time has passed that the accumulated WD proper motions are measurable from second- and third-epoch imaging with sufficient precision to separate cluster and field WDs.   

\citet{2010A&A...513A..50B} use multi-epoch wide-field imaging to measure proper motions for sources in the field of M67 brighter than $V\sim 26$.  They identify cluster members iteratively via their position in the $V, \bv$ color-magnitude diagram (CMD) and their position in the proper motion vector-point diagram, but the Bellini et al.\  analysis focuses on identifying the faintest (oldest) WDs in order to compare the cluster age derived from WD cooling to the age determined by main sequence fitting.  In this paper, we expand the use of Bellini et al.'s proper motion measurements to constrain cluster membership of the brighter, younger WDs in our sample.  This proves to be crucial to producing a clean sample of cluster member WDs and permit studies of the stellar evolutionary topics detailed above.

\subsection{Adopted Cluster Parameters for Messier 67}
For this work we adopt the second-ranked cluster photometric parameters from \citet{2015MNRAS.450.2500B}, which are consistent with those of \citet{2014MNRAS.444.2525C}: $d=880\pm 20$ pc, $E(\bv)=0.02\pm 0.02$, $Z=0.019\pm 0.003$, and age = $3.54\pm 0.15$ Gyr.  This choice was guided in large part by the desire to use a consistent set of isochrones \citep[the PARSEC 1.2S isochrones,][]{2015MNRAS.452.1068C} for the different open cluster WD populations analyzed in \citetalias{PCanton2017}.  As mentioned in Section \ref{sec.intro.mass}, the exact choice of cluster parameters has little impact on our conclusions beyond the precise values of derived WD progenitor masses.  We use the $A/E(\bv)$ determinations from \citet{2011ApJ...737..103S} for the SDSS filters to calculate apparent distance modulus $(m-M)_g = 9.784\pm 0.05$ and color excesses  $E(u-g) = 0.019\pm 0.019$ and $E(g-r)=0.02\pm 0.02$.  The stated errors are obtained by propagating the published uncertainties of \citet{2015MNRAS.450.2500B} through and are treated as Gaussian distributions in magnitude space through the analysis below\footnote{Since the original submission of this paper, GAIA Data Release 2 parallaxes for M67 have been published in \citet{Babusiaux2018}.  Their measured parallax of $\varpi=1.1325\pm0.0011$ mas and $\log(\mathrm{age})=9.54$ are consistent with those we have adopted and do not affect our analysis significantly, so we retain our original age and distance adoptions.}.

\section{Observations}
\subsection{MMT Megacam Imaging}
We obtained imaging data with the Megacam wide-field CCD imager on the MMT 6.5-m telescope over three nights in 2004 and 2005.  Megacam, described fully in \citet{2015PASP..127..366M}, consists of 36 2k$\times$4k CCDs with a scale of 0\farcs08 pix$^{-1}$, resulting in a $25\arcmin\times 25\arcmin$ field of view.  Images were dithered using a five-point pattern to fill in CCD chip gaps.  Total exposure times were 21000 s in $u$, 2400 s in $g$, and 1800 s in $r$.  A summary of the imaging is given in Table \ref{tab.imobslog}.

\begin{deluxetable}{lccc}
\caption{MMT Megacam Imaging Observing Log\label{tab.imobslog}}
\tablehead{\colhead{UT Date} & \colhead{Filter} & \colhead{Exposure Times} & \colhead{FWHM} \\
 & & \colhead{$N_{\rm exp}\times t_{\rm exp}$ (s)} & \colhead{(arcsec)}}
\startdata
2004 March 22 & $g$ & $10\times 240$ & 0.6 \\
2004 March 22 & $r$ & $10\times 180$ & 0.8 \\
2005 January 11 & $u$ & $17\times 600$ & 1.3\\
2005 January 13 & $u$ & $18\times 600$ & 1.3\\
\enddata
\end{deluxetable}

\begin{deluxetable}{CCCc}
\caption{Derived Photometric Coefficients\label{tab.photcoeff}}
\tablehead{\colhead{Filter} & \colhead{Zero Point\tablenotemark{a}} & \colhead{Color Term}  \\
 & \colhead{\rm (mag)} & \colhead{\rm (mag)} }
\startdata
u & a_0 = 24.784\pm 0.012 & a_1 = 0.004\pm 0.006 \\
g & b_0 = 26.159\pm 0.006 & b_1 = -0.079\pm 0.008 \\
r & c_0 = 26.023\pm 0.006 & c_1 = 0.114\pm 0.008 \\
\enddata
\tablenotetext{a}{Nights were not photometric}
\end{deluxetable}

Imaging data are reduced using the 2005 version of the Megacam pipeline described in \citet{2015PASP..127..366M}.  In summary, overscan, and bias corrections are applied separately for each CCD chip in each exposure using the IRAF\footnote{IRAF is distributed by the National Optical Astronomy Observatories, which are operated by the Association of Universities for Research in Astronomy, Inc., under cooperative agreement with the National Science Foundation.} ccdproc task.  Flat fielding is accomplished using twilight flats.  No dark current corrections are applied.  The pipeline applies illumination corrections to each exposure, and astrometric solutions are determined for each frame.  Finally, all exposures are transformed to a tangent plane projection and long exposures in each filter are combined using SWarp \citep{2002ASPC..281..228B}.   The stacked images are qualitatively excellent except for the region around \object{HD 75700}, which fell into chip gaps during some portions of the dither pattern, resulting in a variable scattered light pattern. 

We deliberately choose to perform photometry using aperture photometry.  The field of M67 is not crowded and exhibits minimal blending of sources.  Additionally, PSF photometry from dithered, stacked, and mosaicked images have been previously found to be susceptible to significant errors as compared to aperture photometry on the same images \citep{Slesnick2002}, due in part to abrupt changes in the PSF at chip edges.  Further, our spectroscopy benefits from avoiding blended stars, and our quantitative analysis below is limited much more severely by our spectroscopy than by photometric uncertainties.

We obtain aperture photometry via DAOPHOT II \citep{1987PASP...99..191S}.   We set the aperture radius equal to the measured full width at half maximum (FWHM) in the stacked image of the corresponding filter (see Table \ref{tab.imobslog}).  In order to appear in the final catalog, objects were required to appear in both $g$ and $r$ photometric catalogs.

We exclude extended objects by using two-aperture photometry in the $g$ image with aperture radii of $1\times$ and $2\times {\rm FWHM}$.  Point sources have empirically determined magnitude differences in the two apertures of $ 0.28\pm 0.035$ mag, while extended objects have aperture magnitude differences $\gtrsim 0.5$ mag.  All objects with aperture magnitude differences $\geq 0.42$ mag are therefore considered to be extended and excluded from further analysis.  While this excludes close blends of point sources, the aperture photometry of all such sources is contaminated by the close neighbor and uncontaminated spectroscopy is not easily obtained.

As the images were obtained on non-photometric nights, we calibrate the photometry using photometric data from the SDSS Data Release 12 (DR12) SkyServer \citep{2015ApJS..219...12A}.  Stars with clean SDSS $ugr$ psf photometry are selected from the DR12 catalog and matched with stars in our aperture photometry catalog.  Zero points and color terms are obtained by fitting the following linear functions for all stars with photometric errors $\leq 0.1$ mag:

\begin{eqnarray*}
u_{\rm SDSS} & = & u_{\rm MMT} + 2.5\log_{10} t_{\rm exp} + a_0 + a_1 (u_{\rm SDSS} - g_{\rm SDSS}) \\
g_{\rm SDSS} & = & g_{\rm MMT} + 2.5\log_{10} t_{\rm exp} +b_0 + b_1 (g_{\rm SDSS} - r_{\rm SDSS}) \\
r_{\rm SDSS} & = & r_{\rm MMT} + 2.5\log_{10} t_{\rm exp} +c_0 + c_1 (g_{\rm SDSS} - r_{\rm SDSS}) \\
\end{eqnarray*}

\begin{figure}
\begin{center}
\includegraphics[trim={0 2in 0 1in},clip,width=\columnwidth]{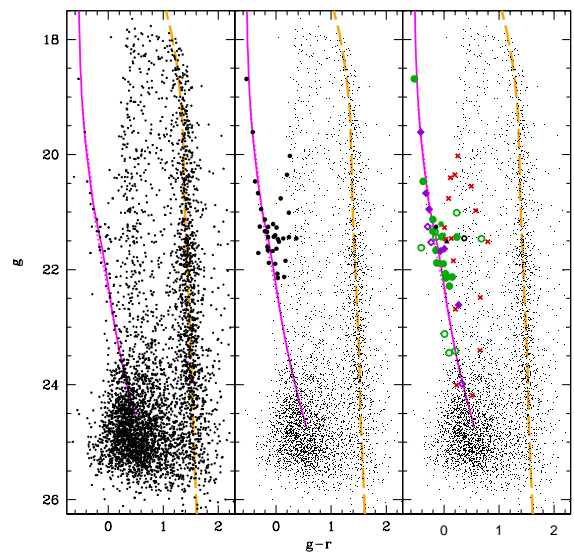}
\end{center}
\caption{The $g-r,g$ color magnitude diagram for the field of M67.  The magenta solid line is a $0.6 M_\sun$ DA WD cooling curve for cooling ages $\leq 4$ Gyr; the magenta dotted line is a $0.6 M_\odot$ DB WD cooling curve for the same age range.  The long-dashed orange line is a PARSEC 1.2S isochrone for a $Z=0.0152$, 3.5 Gyr old stellar population at the adopted distance and extinction.  The left panel shows all unresolved sources with good photometry. The central panel indicates the post-facto photometrically selected WD candidates described in Section \ref{spec.cand_sel} with large points; all other unresolved sources with good photometry are small points. The right panel identifies objects by spectral type.  Green circles are DA WDs, purple diamonds are non-DA WDs, red crosses are non WDs, black circles are photometric targets without spectral identifications, and the black trangle is a photometric target with neither spectral identification nor proper motion data. Filled symbols indicate cluster members, while open symbols indicate non-member objects.  The green asterisks indicate the potential double degenerates. \label{fig.cmds}}
\end{figure}

\begin{figure}
\begin{center}
\includegraphics[trim={0 2in 0 1in},clip,width=\columnwidth]{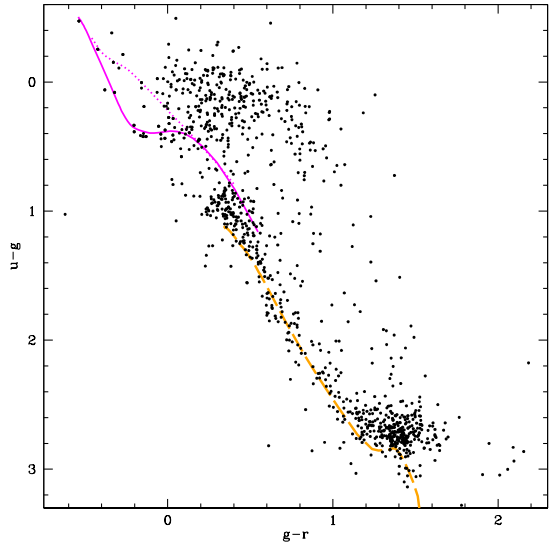}
\end{center}
\caption{The $g-r,u-g$ color color plot for unresolved objects in the field of M67.  The magenta solid line is a $0.6 M_\sun$ DA WD cooling curve for cooling ages $\leq 4$ Gyr; the magenta dotted line is a $0.6 M_\odot$ DB WD cooling curve for the same age range.  The long-dashed orange line is a PARSEC 1.2S isochrone for unevolved stars in a $Z=0.0152$, 3.5 Gyr old stellar population at the adopted distance and extinction.\label{fig.cc}}
\end{figure}

The photometric coefficients are given in Table \ref{tab.photcoeff}.  The residuals to these linear fits exhibit no trends with magnitude, color, or spatial coordinate, though we note that the $u$-band fits are poorly constrained for objects with $(u-g)\leq 0.75$. 
The resulting color-magnitude diagram (CMD) and color-color diagram are presented in Figures \ref{fig.cmds} and \ref{fig.cc}.

\startlongtable
\begin{deluxetable*}{rccccDcDcl}
\tablewidth{0pt}
\tablecaption{White Dwarf Candidate Photometry \label{tab.photcands}}
\tablehead{\colhead{Oblect} & \colhead{RA} & \colhead{Dec} & \dcolhead{g} & \dcolhead{dg} & 
\twocolhead{$u-g$} & \dcolhead{d(u-g)} & \twocolhead{$g-r$} & \dcolhead{d(g-r)} &  \colhead{Cross ID\tablenotemark{b}} \\ 
\colhead{ID\tablenotemark{a}} & \colhead{(J2000)} & \colhead{(J2000)} & \colhead{mag} & \colhead{mag} & \twocolhead{mag} & \colhead{mag} & \twocolhead{mag} & \colhead{mag} &  }
\decimals
\startdata
207325 & 8:50:29.47 & 11:52:49.5  & 20.671 & 0.032 & -0.152 & 0.032 & -0.329 & 0.045 & \objectname[Cl* NGC 2682 MMJ 5011]{MMJ 5011} (1)\\
206310 & 8:50:35.21 & 11:50:33.9  & 20.024 & 0.032 &  0.669 & 0.032 &  0.250 & 0.045 & \objectname[Cl* NGC 2682 MMJ 5055]{MMJ 5055} (1)\\
207569 & 8:50:39.43 & 11:53:26.8  & 21.342 & 0.032 &  0.415 & 0.034 & -0.165 & 0.045 & \\
203383 & 8:50:47.60 & 11:43:30.0  & 21.883 & 0.032 &  0.423 & 0.035 & -0.131 & 0.046 & \\
203493 & 8:50:48.28 & 11:43:47.8  & 21.665 & 0.032 &  0.423 & 0.034 & -0.149 & 0.045 & \object{SDSS J085048.27+114347.7} (7)\\
202728 & 8:50:51.65 & 11:42:06.5  & 21.526 & 0.032 &  0.071 & 0.033 & -0.235 & 0.045 & \\
292232 & 8:50:52.29 & 11:44:43.1  & 21.257 & 0.032 &  0.190 & 0.033 & -0.144 & 0.045 & \object{LB 6310} (3)\\
206986 & 8:50:52.52 & 11:52:06.8  & 21.436 & 0.032 &  0.399 & 0.033 & -0.147 & 0.045 & \\
292440 & 8:50:56.57 & 11:45:21.4  & 20.352 & 0.032 &  0.218 & 0.032 &  0.198 & 0.045 & \object{LB 6316} (3)\\ 
204474 & 8:50:57.45 & 11:46:08.4  & 21.471 & 0.032 &  0.283 & 0.034 &  0.033 & 0.045 & \\
204254 & 8:50:58.61 & 11:45:38.8  & 22.127 & 0.032 &  0.350 & 0.037 &  0.147 & 0.046 & \objectname[\[VTB2004\] CX 86]{CX 86} (2)\\
205452 & 8:51:01.61 & 11:48:26.0  & 21.253 & 0.032 & -0.108 & 0.032 & -0.296 & 0.045 & \object{LB 6319} (3)\\
207230 & 8:51:01.78 & 11:52:34.5  & 21.620 & 0.032 &  0.162 & 0.033 & -0.413 & 0.046 & \\
203546 & 8:51:05.31 & 11:43:56.7  & 21.333 & 0.032 &  0.386 & 0.033 & -0.203 & 0.045 & \\
8885     & 8:51:08.89 & 11:45:44.9  & 21.404 & 0.022 &  0.372 & 0.025 & -0.016 & 0.032 & \object{LB 6326} (3) \\
7105     & 8:51:09.58 & 11:43:52.6  & 21.011 & 0.022 &  0.498 & 0.024 &  0.231 & 0.032 & \object{LB 6328} (3) \\
15952   & 8:51:12.11 & 11:52:31.3  & 21.129 & 0.024 &  0.335 & 0.025 & -0.204 & 0.040 & \object{LB 6329} (3) \\
17303   & 8:51:18.78 & 11:53:49.7  & 21.458 & 0.023 &  0.484 & 0.024 &  0.367 & 0.039 & \objectname[\[VTB2004\] CX 64]{CX 64} (2)\\
12051   & 8:51:19.90 & 11:48:40.6  & 18.685 & 0.032 & -0.472 & 0.032 & -0.538 & 0.045 & \object{LB 6339} (3,4)\\
18288   & 8:51:21.25 & 11:54:44.6  & 22.145 & 0.032 &  0.412 & 0.035 &  0.031 & 0.046 & \\
14108   & 8:51:21.33 & 11:50:43.2  & 21.430 & 0.032 &  0.404 & 0.033 & -0.042 & 0.045 & \object{LB 6343} (3) \\
15645   & 8:51:21.61 & 11:52:08.9  & 21.847 & 0.032 &  0.191 & 0.034 &  0.170 & 0.045 & \objectname[\[VTB2004\] CX 89]{CX 89} (2)  \\
16860   & 8:51:22.64 & 11:53:29.2  & 21.677 & 0.032 & -0.026 & 0.033 & -0.063 & 0.045 & \\
17469   & 8:51:24.92 & 11:53:56.6  & 21.209 & 0.032 &  0.343 & 0.034 & -0.055 & 0.046 & \\
3477     & 8:51:32.14 & 11:39:20.3  & 20.762 & 0.032 &  0.578 & 0.032 &  0.083 & 0.045 & \object{SDSS J085132.12+113920.4} (8)  \\
5827     & 8:51:34.88 & 11:42:17.7  & 19.610 & 0.032 & -0.254 & 0.032 & -0.424 & 0.045 & \object{LB 3600} (3,5)\\
21761   & 8:51:37.75 & 11:58:43.2  & 21.436 & 0.033 &  0.331 & 0.037 &  0.234 & 0.051 & \\
9250     & 8:51:40.56 & 11:46:00.6  & 20.468 & 0.032 &  0.061 & 0.032 & -0.381 & 0.045 & \object{LB 3601} (3) \\
4520     & 8:51:40.96 & 11:40:30.5  & 22.073 & 0.032 &  0.461 & 0.035 &  0.024 & 0.046 & \\
4941     & 8:51:45.20 & 11:41:04.6  & 21.900 & 0.032 &  0.436 & 0.034 & -0.017 & 0.046 & \object{LB 6373} (3)\\
11913   & 8:51:45.72 & 11:48:34.6  & 20.949 & 0.032 & -0.213 & 0.032 & -0.271 & 0.045 & \\
15555   & 8:51:46.56 & 11:52:02.6  & 21.459 & 0.032 &  0.393 & 0.033 &  0.131 & 0.045 & \object{LB 6374},\objectname[\[VTB2004\] CX 33]{CX 33} (2,3) \\
5409     & 8:51:48.93 & 11:41:43.4  & 21.638 & 0.032 &  0.275 & 0.034 &  0.006 & 0.045 &  \\
15623   & 8:52:07.35 & 11:52:07.3  & 21.515 & 0.032 &  0.061 & 0.033 &  0.051 & 0.046 &  \\
5072     & 8:52:07.47 & 11:41:14.5  & 21.268 & 0.032 &  0.322 & 0.033 &  0.008 & 0.045 &  \\
\midrule
208218 & 8:50:49.27 & 11:54:59.1  & 21.516 & 0.032 &  1.095 & 0.040 &  0.794 & 0.045 & \\
201033 & 8:51:06.21 & 11:37:54.1  & 20.404 & 0.032 & -0.026 & 0.032 &  0.116 & 0.045 & \object{LB 3598} (3) \\
8358     & 8:51:09.14 & 11:45:20.2  & 21.464 & 0.023 &  0.449 & 0.025 &  0.680 & 0.032 & \object{LB 6327} (3) \\
11531   & 8:51:11.04 & 11:48:14.3  & 23.119 & 0.023 &  0.442 & 0.040 &  0.009 & 0.035 & \\
20808   & 8:51:29.95 & 11:57:33.0  & 22.285 & 0.032 &  0.434 & 0.039 &  0.099 & 0.046 & \\
3784     & 8:51:38.43 & 11:39:41.8  & 23.986 & 0.039 &  0.425 & 0.097 &  0.327 & 0.061 & \\
11379   & 8:51:39.75 & 11:48:05.1  & 23.413 & 0.035 &  0.406 & 0.062 &  0.205 & 0.054 & \\
15202   & 8:51:43.56 & 11:51:43.6  & 22.615 & 0.033 &  0.372 & 0.041 &  0.266 & 0.048 & \\
4368     & 8:51:44.36 & 11:40:22.0  & 20.975 & 0.032 &  0.198 & 0.032 &  0.573 & 0.045 & \\
10197   & 8:51:45.02 & 11:46:56.4  & 23.447 & 0.038 &  0.434 & 0.073 &  0.091 & 0.073 & \\
3495     & 8:51:49.14 & 11:39:21.8  & 20.552 & 0.032 &  1.136 & 0.033 &  0.493 & 0.045 & \objectname[Cl* NGC 2682 MMJ 6157]{MMJ 6157} (1) \\
13048   & 8:51:51.21 & 11:49:37.3  & 24.186 & 0.041 &  0.508 & 0.130 &  0.515 & 0.065 & \\
10924   & 8:51:52.19 & 11:47:39.0  & 24.006 & 0.039 &  0.546 & 0.110 &  0.226 & 0.066 & \\
14428   & 8:51:57.72 & 11:51:04.9  & 22.697 & 0.032 &  0.506 & 0.045 &  0.206 & 0.047 & \\
4432     & 8:52:08.40 & 11:40:25.8  & 23.405 & 0.036 &  0.251 & 0.055 &  0.653 & 0.052 &  \\
4888     & 8:52:09.34 & 11:40:59.4  & 22.484 & 0.033 &  0.760 & 0.044 &  0.659 & 0.046 &  \\
2231     & 8:52:12.00 & 11:37:50.4  & 19.114 & 0.032 &  0.473 & 0.046 &  0.190\tablenotemark{c} & 0.024\tablenotemark{c} & \objectname[Cl* NGC 2682 FBC 4895]{FBC 4895} (6) \\
\enddata
\tablecomments{Entries below the bar do not meet the \emph{post facto} photometric criteria but were targeted spectroscopically  either due to results of first-iteration photometric reduction or the desire to fill slit masks with targets to explore parameter space outside the formal selection criteria.}
\tablenotetext{a}{Running index from DAOphot photometry files used as identifier in slit mask design}
\tablenotetext{b}{Simbad coordinate matches within 3\arcsec radius}
\tablenotetext{c}{Star off plate in our $r$ image; data from SDSS DR14}
\tablerefs{(1) \citet{1993AJ....106..181M}; (2) \citet{2004AA...418..509V}; (3) \citet{1963LB....C32....1L};  (4) \citet{1994AA...290L..17P}; (5) \citet{1997ASSL..214...91F}; (6) \citet{1996AJ....112..628F}; (7) \citet{Kepler2015}; (8) \citet{2014AA...563A..54P} }
\end{deluxetable*}

\subsection{Keck Spectroscopy}
\subsubsection{Candidate White Dwarf Selection \label{spec.cand_sel}}

We select spectroscopic targets guided by a comparison of photometric data with predicted WD cooling models while avoiding portions of color space inhabited by main sequence stars.  The selection region spans the most common masses and spectral types of WDs while excluding thick disk MS turnoff stars and unresolved blue galaxies. A handful of objects with blue $u-g$ colors but red $g-r$ colors are also selected as potential WD+M binary systems.  We select spectroscopic targets targets without consideration of proper motion memberships, as these memberships were not available to us prior to spectroscopic observations.

Our initial target selection criteria were defined prior to finalization of photometric calibrations and did not apply a limiting magnitude.  Guided by subsequent evolution of the photometric calibrations and the need for high signal-to-noise spectral data, we define post factum quantitative photometric criteria for WD candidates: all unresolved objects with $g\leq 22.2$, $-0.6\leq u-g \leq 0.7$, and $ 1.05+2.05(g-r)\leq u-g \leq -0.03 + 1.214(g-r)$ are considered to be candidate WDs.  

\begin{figure}
\begin{center}
\includegraphics[trim={0 2in 0 1in},clip,width=\columnwidth]{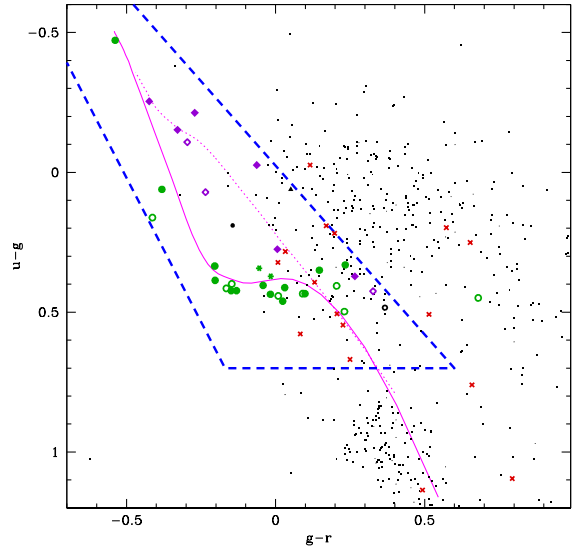}
\end{center}
\caption{A close-up of the WD region of the $g-r,u-g$ color-color plot for the field of M67.  The magenta solid line is a $0.6 M_\sun$ DA WD cooling curve for cooling ages $\leq 4$ Gyr; the magenta dotted line is a $0.6 M_\odot$ DB WD cooling curve for the same age range.   Tiny points are all unresolved point sources with good photometry.  Large symbols identify objects by spectral identifications:  green circles are DA WDs, purple diamonds are non-DA WDs, red crosses are non WDs, the black circles are photometric WD candidates without spectra, and the black triangle is a photometric WD candidate with neither a spectrum nor a proper motion measurement. Filled symbols indicate cluster members, while open symbols indicate non-member objects.  The green asterisks indicate the potential double degenerates. \label{fig.cc_cands}}
\end{figure}

This post facto selection region is shown in the color-color diagram in Figure \ref{fig.cc_cands}, which presents the color-color diagram for bluer point sources in the MMT imaging, along with representative 0.6\msun DA and DB WD cooling curves for cooling ages less than 4 Gyr. Positional and photometric data for the  candidates thus selected are given in Table \ref{tab.photcands} and are  plotted in the center panel of Figure \ref{fig.cmds}.  Data for the 17 spectroscopically targeted sources not meeting the post facto selection criteria are given below the bar in Table \ref{tab.photcands}.

Spectroscopic target selection was also influenced by slit mask placement and orientation.  These were determined by qualitatively maximizing the number of highest priority targets (i.e., candidate WDs selected by the photometric criteria) that could fit on a slitmask.  The primary selection bias for M67 member WDs is therefore based on luminosity, with potential but likely minimal secondary biases due to binarity and any differences in spatial distributions of WD subpopulations.  

\subsubsection{Observing details}

We obtained spectroscopic data with the Low-Resolution Imaging Spectrograph \citep[LRIS,][]{1995PASP..107..375O,1998SPIE.3355...81M,2004ApJ...604..534S,2010SPIE.7735E..0RR} on the Keck I telescope over four nights in 2007 and 2010. A run in 2006 February was lost to thick clouds.  We obtained most spectra using slitmasks with 1\arcsec~wide slitlets, though we used a 1\arcsec~wide long slit for spectrophotometric calibrators and for \objectname{LB 3600}.  Although we collected data using both red and blue arms of LRIS, only the blue arm data are presented here.  An observing log for the 2007 and 2010 runs is presented in Table \ref{tab.specobslog}.

\begin{deluxetable*}{lcccl}
\caption{Spectroscopic Observing Log\label{tab.specobslog}}
\tablehead{\colhead{UT Date} & \colhead{Mask} & \colhead{\# exp} & \colhead {exp time} & \colhead{Comments} \\ & & & (s) & }
\startdata
2007 Jan 19 & f2 & 9 & 1200 & FWHM$=$0\farcs9  \\
                     & f3 & 6 & 1200 & \\
2007 Jan 20 & f1 & 9 & 1200 & FWHM$=$1\farcs0; clear skies \\
                     & f4 & 8 & 1200 & \\
                     & LS\tablenotemark{a} & 2 & 600 & Target: LB 3600\\
2010 Feb 8   & ma & 9 & 1200 & FWHM$= 1\arcsec$\\
                     & md & 10 & 1200 & Seeing steadily degrading; FWHM$=$1\arcsec to 1\farcs5\\
2010 Feb 9   & mc & 6 & 1200 & \\
                     & mb & 3 & 1200 & Poor guiding \\
                     & me & 9 & 1200 & \\
\enddata
\tablenotetext{a}{1\farcs 0 wide longslit}
\end{deluxetable*}

The spectroscopic configuration used the 400/3400 grism and D560 dichroic. The 400/3400 grism was selected over the higher-resolution 600/4000 grism due to its higher throughput shortward of 4000\AA, where the mass-sensitive higher order Balmer lines are located.  Resulting spectral resolutions, as measured by the FWHM of the 5577\AA~night sky emission line, are $\approx 8.0$ \AA, with some variation depending on the slitlet location.

We designed slitmasks using the Autoslit3 program provided by J.~Cohen.  Slitmask position and orientations were varied to permit as many WD candidates as possible to fit on a slit mask; typically three to five high priority candidates were present on a slitmask.  If WD candidate slit placement was not affected, additional slitlets were milled for serendipitous studies of other objects, such as galaxies and unresolved objects outside the stellar locus in the color-color diagram.  Data from these extra slitlets are not presented in this paper.  The Keck I atmospheric dispersion corrector was in use and operational, so little heed was paid to parallactic angle.  A pickoff mirror was attached to each slitmask for purposes of guiding.

\subsubsection{Spectroscopic data reduction}

\begin{figure*}
\includegraphics[scale=0.67,angle=270]{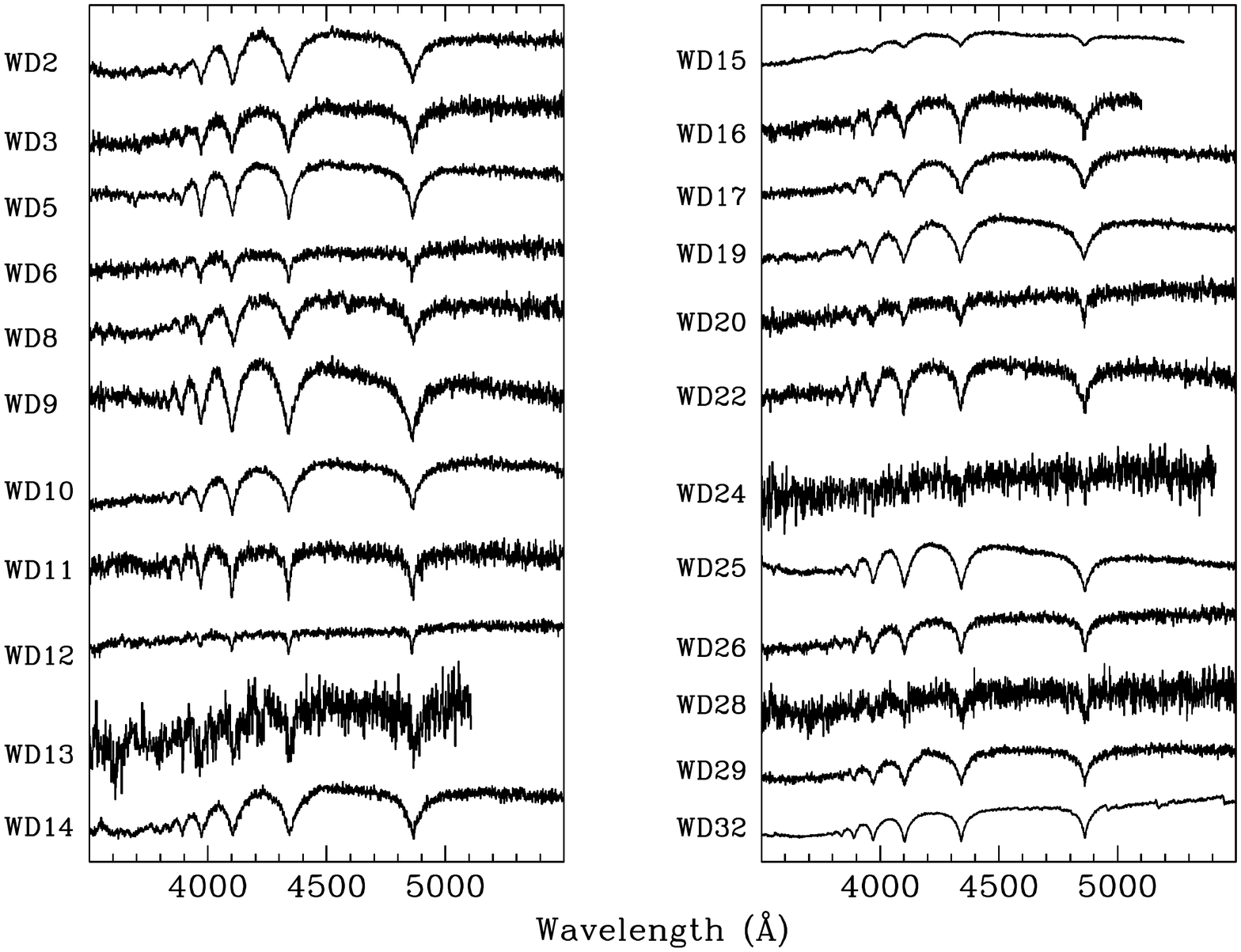}
\caption{LRIS-B spectra for DA WDs in the field of M67.  The spectra are calibrated in $f_\nu$ units and normalized to the flux at 5500\AA~or the reddest observed wavelength, with an arbitrary vertical offset applied. \label{fig.daspec}}
\end{figure*}

\begin{deluxetable}{rccccc}
\tablecaption{Spectroscopically Confirmed White Dwarfs \label{tab.speccands}}
\tablehead{\colhead{Object} & \colhead{Alias} & \dcolhead{g} & \dcolhead{dg} & \colhead{Mask\tablenotemark{b}} & \colhead{Spectral} \\ 
ID\tablenotemark{a} & & \colhead{mag} & \colhead{mag} & & \colhead{Type} }
\startdata
207325 & WD1 & 20.671 & 0.032 & ma & DB \\
207569 & WD2 & 21.342 & 0.032 & ma & DA \\
203383 & WD3 & 21.883 & 0.032 & f2 & DA \\
202728 & WD4 & 21.526 & 0.032 & f2 & DB \\
206986 & WD5 & 21.436 & 0.032 & ma & DA \\
204254 & WD6 & 22.127 & 0.032 & f2 & DA \\
205452 & WD7 & 21.253 & 0.032 & md & DB \\
207230 & WD8 & 21.620 & 0.032 & md & DA \\
203546 & WD9 & 21.333 & 0.032 & f1 & DA \\
8885     & WD10 & 21.404 & 0.022 & f2 & DA \\
8358     & WD11 & 21.464 & 0.023 & f1 & DA \\
7105     & WD12 & 21.011 & 0.022 & f1 & DA  \\
11531   & WD13 & 23.119 & 0.023 & md & DA \\
15952  & WD14 & 21.129 & 0.024 & md & DA \\
12051  & WD15 & 18.685 & 0.032 & f2 & DA \\
18288  & WD16 & 22.145 & 0.032 & f4 & DA \\
14108  & WD17 & 21.430 & 0.032 & f3 & DA \\
16860  & WD18 & 21.677 & 0.032 & f4 & DB \\
17469  & WD19 & 21.209 & 0.032 & f4 & DA \\
20808  & WD20 & 22.285 & 0.032 & f4 & DA \\
5827    & WD21 & 19.610 & 0.032 & LS & DB \\
21761  & WD22 & 21.436 & 0.033 & f4 & DA \\
3784    & WD23 & 23.986 & 0.039 & me & DB: \\
11379  & WD24 & 23.413 & 0.035 & mc & DA \\
9250    & WD25 & 20.468 & 0.032 & mc & DA \\
4520    & WD26 & 22.073 & 0.032 & me & DA \\
15202  & WD27 & 22.615 & 0.033 & mc & DC \\
10197  & WD28 & 23.447 & 0.038 & mc & DA \\
4941    & WD29 & 21.900 & 0.032 & me & DA \\
11913  & WD30 & 20.949 & 0.032 & mc & DB \\
5409    & WD31 & 21.638 & 0.032 & me & DC \\
2231    & WD32 & 19.114 & 0.032 & mb & DA+dM \\
203493 &WD33 & 21.665 & 0.032 & \nodata & DA\tablenotemark{c} \\
\enddata
\tablenotetext{a}{Running index from DAOphot photometry files used as identifier in slit mask design}
\tablenotetext{b}{As given in Table \ref{tab.specobslog}}
\tablenotetext{c}{Spectral type from \citet{Kepler2015}}
\end{deluxetable}

\begin{figure}
\begin{center}
\includegraphics[trim={0 2in 0 1in},clip,width=\columnwidth]{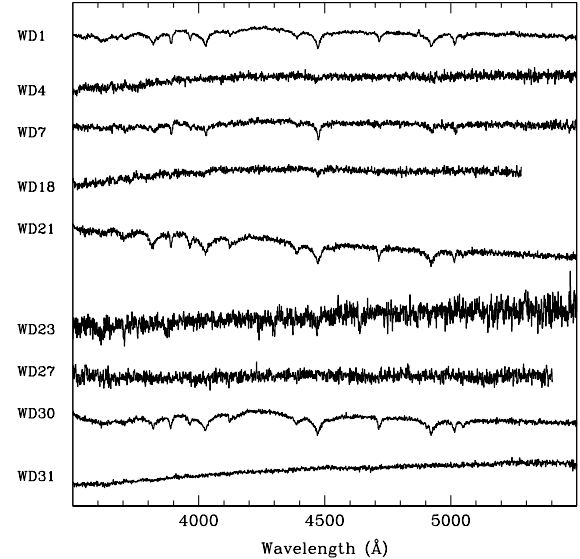}
\end{center}
\caption{LRIS-B spectra for non-DA WDs in the field of M67.  The spectra are calibrated in $f_\nu$ units and normalized to the flux at 5500\AA, with an arbitrary vertical offset applied. \label{fig.nondaspec}}
\end{figure}

\begin{figure*}
\begin{center}
\includegraphics[scale=0.67]{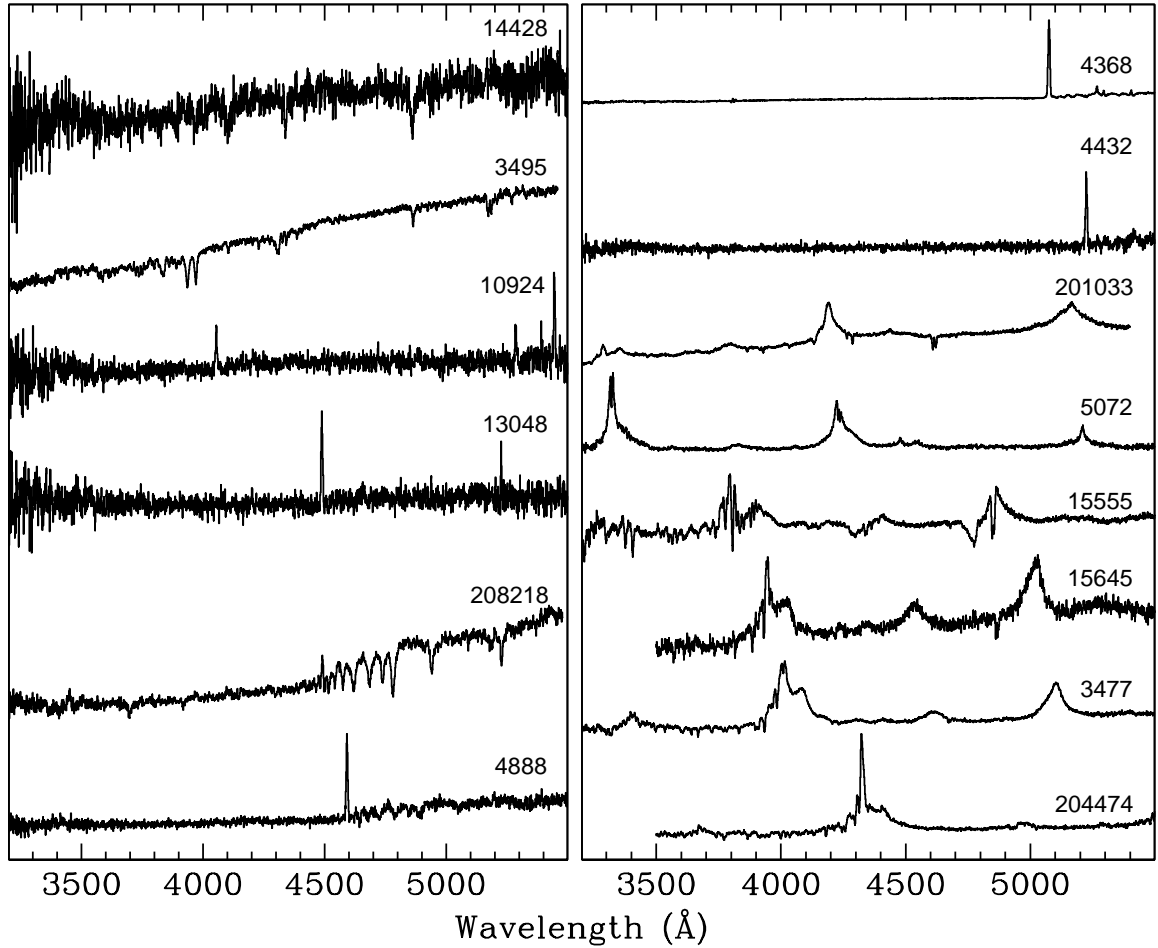}
\end{center}
\caption{LRIS-B spectra for non WDs in the field of M67.  The spectra are calibrated in $f_\nu$ units.  Absorption-line objects are normalized to the flux at 5500\AA, with an arbitrary vertical offset applied. Emission-line spectra are normalized to the peak of the highest emission feature.  Spectra are ordered by increasing redshift. \label{fig.nonwdspec}}
\end{figure*}

\begin{deluxetable}{rcclC}
\tablecaption{Spectroscopic Identifications of Non-White Dwarfs \label{tab.nonwds}}
\tablehead{\colhead{Object ID} & \dcolhead{g} & \dcolhead{dg} & \colhead{Spectral} & \colhead{z} \\ 
 & \colhead{mag} & \colhead{mag} & \colhead{ID} & }
\startdata
206310\tablenotemark{c} & 20.024 & 0.032 & QSO\tablenotemark{c} & 2.555\tablenotemark{c} \\
208218 & 21.516 & 0.032 & E+A & 0.205 \\
292440\tablenotemark{c} & 20.352 & 0.032 & QSO\tablenotemark{c} & 1.184\tablenotemark{c} \\ 
204474 & 21.471 & 0.032 & QSO & 2.56 \\
201033 & 20.404 & 0.032 & QSO & 1.70 \\
15645  & 21.847 & 0.032 & QSO & 2.24 \\
3477    & 20.762 & 0.032 & QSO & 2.29 \\
4368    & 20.975 & 0.032 & ELG\tablenotemark{b} & 0.36 \\
15555  & 21.459 & 0.032 & QSO & 2.14 \\
3495    & 20.552 & 0.032 & F/G star & \nodata \\
13048  & 24.186 & 0.041 & ELG\tablenotemark{b} & 0.20 \\
10924  & 24.006 & 0.039 & ELG\tablenotemark{b} & 0.088 \\
14428  & 22.697 & 0.032 & A star & \nodata \\
5072    & 21.268 & 0.032 & QSO & 1.73 \\
4432    & 23.405 & 0.036 & ELG\tablenotemark{b} & 0.40 \\
4888    & 22.484 & 0.033 & E+A & 0.23 \\
\enddata
\tablenotetext{a}{Running index from DAOPHOT photometry files used as identifier in slit mask design}
\tablenotetext{b}{Emission-line galaxy}
\tablenotetext{c}{Not observed in this work; data from \citet{2014AA...563A..54P}}
\end{deluxetable}

We reduced the spectroscopic data using standard techniques in the IRAF twodspec and onedspec packages. We used overscan regions to subtract amplifier bias from individual images; we then used the L.A. Cosmic cosmic-ray rejection routine \citep{2001PASP..113.1420V} to flag and remove cosmic rays from the two-dimensional spectra.  We median combined internal flat field exposures for each slitmask and normalized these via a polynomial fit in the spectral direction over wavelengths where the flat field counts were higher than 300 cts, corresponding to wavelengths $\gtrsim 3600$\AA.  Residuals to the polynomial fit had an RMS $\leq 0.5\%$.  The normalized spectroscopic flat fields were applied to each individual exposure.  

Spectrograph flexure in the spatial direction was significant.  We determined the centroids of the spectroscopic traces of mask alignment stars to determine individual shifts in the spatial direction for each exposure prior to combining the exposures.  In order to avoid interpolation issues, we rounded these shifts to the nearest integer number of pixels and shifted each image in the spatial direction by this amount.  Since the spatial scale of the blue arm of LRIS is $ 0\,\farcs 135\, {\rm pix}^{-1}$ and typical seeing was $\approx 1\arcsec$, any distortions in the spatial direction are minimal.

We combined the spectra by averaging the resulting images, masking pixels flagged as cosmic rays by L.A.~Cosmic, to produce a single deep spectroscopic exposure for each slitmask. We then extracted 1-dimensional spectra from each slitlet and applied a wavelength solution derived from Hg, Cd and Zn arclamp spectra.  A zeropoint wavelength offset was applied to the arclamp solution to ensure that the centroid of the peak of the [\ion{O}{1}] auroral line was 5577.338\AA~ \citep{1996PASP..108..277O}.  We used 1\arcsec~longslit spectra of \object{G 191-B2B} obtained with the same spectral setup to determine a spectral response function; we applied this to our extracted slitlet spectra to achieve relative flux calibration.  We note that some objects have spectral slopes inconsistent with their photometric colors, most commonly with apparently missing blue flux (e.g., see WD15 in Figure \ref{fig.daspec} for an egregious example).  The cause of this is unclear but is almost certainly due to instrumentation or human error.  Caution should be exercised in interpretation of the spectral slopes.

\subsubsection{Spectroscopic completeness}

Of the 35 photometric WD candidates in the our post facto candidate sample, 29 had successful spectroscopic observations for an 83\% success rate.  An additional 17 objects not meeting the post facto selection criteria resulted in successful spectroscopic data  as well.
We consider a ``successful'' spectroscopic observation of a target to have resulted in sufficient signal-to-noise to permit spectral identification.  WDs exhibiting the hydrogen Balmer line series are identified as spectral type DA; WDs clearly exhibiting at least one absorption line of \ion {He}{1} are identified as DB, and WDs with no obvious spectral features are identified as DC.  Because our spectra do not cover H$\alpha$, it is possible that one or more of our DBs and DCs have H$\alpha$ absorption.  One faint DB, WD23, appears to have additional significant absorption features in addition to \ion{He}{1}~$\lambda 4471$\AA, but the wavelengths of these features do not precisely match any common strong stellar absorption features.  We therefore identify this star as a ``DB:'' following the revised spectral classification scheme presented in \citet{1999ApJS..121....1M} and references therein.

Of the six photometric candidates for which we were unable to obtain successful spectroscopic observations, three have identifications published in the literature.  Objects 206310 and 202440 are both listed in the DR10 version of the SDSS quasar catalog \citep{2014AA...563A..54P}; visual inspection of the SDSS spectra for both objects confirms they are indeed QSOs.  Object 203493 (WD33) is a DA WD in the  DR10 WD catalog of \citet{Kepler2015}, though the signal to noise of the SDSS spectrum is very low. Therefore, only three unresolved objects meeting the post facto candidate selection criteria lack spectral identification: Objects 292232, 17303\footnote{Object ID 17303 is coincident with an X-ray source \citep{2004AA...418..509V} and is therefore likely an AGN, but there is no optical confirmation.}, and 15623, for an incompleteness of 9\%. 

We do not quantify the completeness of the photometric catalog, but as the MMT images are very deep and are not crowded, and as our spectroscopic magnitude limit ($g\leq 22.2$) is well above the photometric limit of $g\lesssim 25.25$, we do not expect any significant systematics to be introduced by photometric incompleteness.

We also expect no systematics in completeness due to atmospheric composition.  The selection criteria include the cooling tracks for both DA and non-DA WDs.  In addition, although WDs with H-dominated and He-dominated atmospheres cool at different rates over different magnitude ranges, this selection effect is not important for our selection range.  Taking $g=22.2\,(M_g\approx 12.4)$ as our magnitude limit and using the $0.6 M_\sun$ evolutionary models of \citet{2001PASP..113..409F}, we find that the cooling time of a DA at our magnitude limit is 698 Myr, while the cooling time for a DB of this magnitude is 696 Myr, nearly identical.

Spectral identifications are presented in Table \ref{tab.speccands} for WDs and in Table \ref{tab.nonwds} for non-WDs.  Spectra for DA WDs are plotted in Figure \ref{fig.daspec}, non-DA WD spectra are plotted in Figure \ref{fig.nondaspec}, and spectra for non-WDs are plotted in Figure \ref{fig.nonwdspec}.   These identifications are also illustrated in the right-hand panels of the cluster CMD and WD color-color plots (Figures \ref{fig.cmds} and \ref{fig.cc_cands}, respectively).

\subsubsection{WD atmospheric parameters}

Atmospheric parameters for DA WDs are determined using techniques described in detail in \citetalias{PCanton2017}.  In short, we fit normalized Balmer line profiles from one dimensional atmospheric models using the methodology and software of \citet{Gianninas2011} to obtain \teff and \logg; their methods are refinements of those presented by \citet{Bergeron1992} and \citet{1995ApJ...449..258B}.  We note that the uncertainties in the fitted parameters include the external errors estimated by \citet{2005ApJS..156...47L}, though using the corrected values of 1.4\% in \teff and 0.042 dex in \logg as discussed in \citet{2017arXiv170902324B}.  We then apply the 3-D \teff and \logg corrections from Appendix C of  \citet{2013A&A...552A..13T} to those WDs with $\teff\leq 15000$ K to account for the known shortcomings mixing length theory produces in one dimensional DA WD atmospheric models.

We obtain synthetic photometry for each DA WD for comparison to observed photometry for the purposes of identifying potential binary systems and cluster membership.  Synthetic photometry is calculated using tabulated color tables (2016 October vintage) provided by P.~Bergeron at \url{http://www.astro.umontreal.ca/\~bergeron/CoolingModels}.  These evolutionary models and color calculations are detailed in \citet{2006AJ....132.1221H,2006ApJ...651L.137K,2011ApJ...730..128T,2011ApJ...737...28B}.  We calculate uncertainties in the synthetic photometry by simulating 1000 observations of the WD with \teff and \logg drawn from a Gaussian distribution centered on the derived \teff and \logg with stated errors, propagating these through all photometric calculations, and then deriving the standard deviation of the output synthetic photometry.

An apparent distance modulus for each DA WD is determined by subtracting the $g$-band apparent magnitude from the synthetic absolute $g$ magnitude.  Color excesses are also calculated by subtracting synthetic colors from the observed colors.  Errors on the apparent distance moduli and color excesses are determined by adding observational uncertainties, cluster distance and reddening uncertainties, and synthetic model uncertainties in quadrature.

\section{The White Dwarf Population of Messier 67}
\subsection{Cluster membership determination}
We determine cluster membership for each WD as follows.  For DA WDs with proper motion measurements, we require the WD to be a proper motion cluster member as defined by \citet{2010A&A...513A..50B} and to have an apparent distance modulus within three standard deviations of our adopted value.  18 of the 24 (75\%)  spectroscopically confirmed DA WDs have spectra of sufficient quality to permit photometric membership determination. A search of the Gaia DR2 catalog is unable to better constrain cluster membership -- none of our WDs have sufficiently precise parallax measurements, and only WD 15 and WD 21 have significant proper motions, both of which are consistent with cluster membership in both the Gaia and \citet{2010A&A...513A..50B} catalogs.

Candidate double degenerates are identified as DA WD proper motion members with distance moduli inconsistent with cluster membership yet foreground to the cluster up to a maximum of 0.75 mag (i.e., those less than a factor of two overluminous). We note that the spectrum of any true double degenerate system will be a flux-averaged combination of the two components, and that these two components need not be the same mass or luminosity.  Therefore, the derived atmospheric parameters of candidate double degenerates should be considered unreliable.  

To date we have not attempted to determine atmospheric parameters for non-DA WDs; we hope to revisit this issue in a future study.  For now, non-DA cluster membership is based solely on the WD's proper motion membership from \citet{2010A&A...513A..50B}.  The information relevant to membership determination is presented in Table \ref{tab.damems} for DA WDs and in Table \ref{tab.dbmems} for non-DA WDs.   

\begin{figure}
\begin{center}
\includegraphics[trim={0 2in 0 1in},clip,width=\columnwidth]{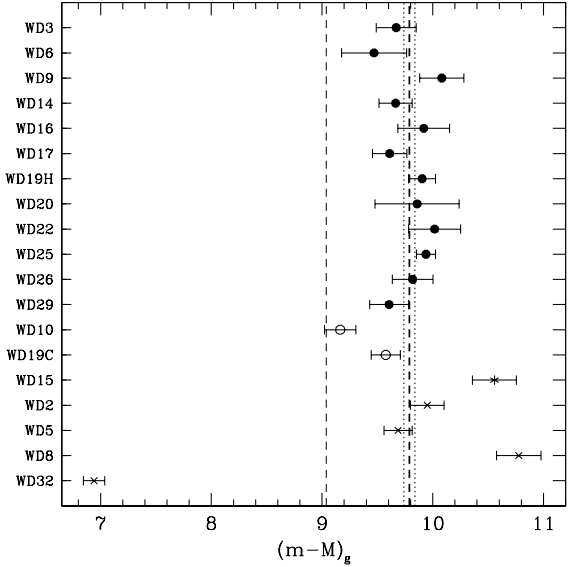}
\end{center}
\caption{Apparent distance moduli for DA WDs in the field of M67.  Filled points are WDs that meet both proper motion and distance modulus criteria for membership; the cluster distance modulus is indicated by the vertical heavy dashed line with uncertainties of the vertical dotted lines.  Open circles are WDs that are proper motion members but up to 0.75 mag foreground to the cluster (indicated by the light dashed vertical line); these are double degenerate candidates.  WD15, indicated by an asterisk, is a proper motion member but significantly background to the cluster.  The four DAs indicated by a cross are non-members according to proper motions.  The two possible solutions for WD19 are shown as WD19H (hotter solution) and WD19C (cooler solution).  Error bars indicate $2\sigma$ uncertainty.  The combination of proper motion and distance modulus criteria identifies likely field WDs better than either criterion alone.   \label{fig.dmods}}
\end{figure}

In Figure \ref{fig.dmods}, we plot the apparent distance modulus for each DA WD as compared to the cluster distance modulus.  This figure highlights the utility of combining photometric and proper motion membership determinations.  Fully half of the nonmembers (based on proper motions) would have been considered cluster member WDs based on distance modulus criteria alone, and the resulting contamination would have been $\sim30\%$.  Additionally, there is one DA WD, WD15, that has a proper motion consistent with M67 membership but a distance modulus that is significantly background to the star cluster.  Such an interloper is not surprising due to the overlap of the field star and cluster star proper motion vectors as seen in Figure 4d of \citet{2010A&A...513A..50B}.  The combination of distance modulus and proper motion measurements is therefore a powerful tool in producing an uncontaminated sample of cluster member WDs.

Of the six post facto photometric WD candidates without new spectroscopy, three are proper motion nonmembers (Objects 206310, 292440, and 17303); two are proper motion members (the DA WD33 and Object 292232); and one, Object 15623, is outside the footprint of the \citet{2010A&A...513A..50B} proper motion data.

\emph{Notes on individual objects:}

\emph{WD10} --- This WD is a proper motion member and candidate double degenerate, as it is otherwise overluminous for a cluster member WD with the derived atmospheric parameters.  However, if a binary, its spectrum would be a composite; our spectral analysis assumes it is a single star.  We therefore exclude it from the cluster WD mass distribution in Section \ref{sec.wdmass}.

\emph{WD15} --- This DA is a proper motion member, but its distance modulus is 0.77 mag ($\approx 8 \sigma$) background to the cluster, which would imply it is a field star.  This would not be unexpected, since the cluster member proper motion distribution in \citet{2010A&A...513A..50B} does overlap with the field proper motion distribution, though to date we have not calculated proper motion membership probabilities.  Additionally, even though the spectral signal-to-noise is high, the Balmer line fits are mediocre for this star, meaning that the derived parameters (and therefore distance modulus) could be in error.  Because of these ambiguities, we do not include it in our further analysis of cluster members.

\emph{WD19} --- This DA is a proper motion member, but two different \teff and \logg solutions are equally good matches to the spectroscopic and photometric data \citepalias[see][]{PCanton2017}.  Both spectral solutions have synthetic photometry consistent with cluster membership, though the cooler solution is also consistent with a potential double degenerate solution.  The derived WD masses are consistent each other to within $2\sigma$.  We therefore consider this WD a cluster member, but we exclude it from the WD mass distribution discussed below.

\emph{WD24} --- This faint DA is a proper motion member, but the low signal-to-noise spectrum does not permit precise determinations of mass or synthetic photometry.  We therefore exclude this WD from further discussion.

\emph{WD28} --- This DA does not have a proper motion measurement from \citet{2010A&A...513A..50B}.  Photometrically, it is 1.22 mag background to the cluster, but the signal-to-noise of the spectrum is very low, resulting in poor constraints on the synthetic photometry.  We exclude this WD in all further analysis.

\emph{WD32} --- This DA shows unambiguous spectroscopic evidence of an unresolved M-star companion redward of the H$\beta$ line.  Given that the WD is not a proper motion member and is significantly foreground to the cluster, we do not analyze the spectrum any further.

\emph{WD33} --- This DA was not observed spectroscopically by us but has spectral data from the SDSS DR10 WD catalog of \citet{Kepler2015}, and it is a proper motion member of the cluster.  \citet{Kepler2015} derive spectral parameters of $\teff=9914 \pm 284$ K and $\logg=8.310\pm 0.300$ using the spectral models of \citet{Koester2010} and 3D corrections, which generally results in parameters consistent with published results from our methodology \citepalias[see][and references therein]{PCanton2017}.  Due to the very low signal-to-noise of the SDSS spectrum (S/N$=3$), the mass and the synthetic photometry are poorly constrained.  We therefore exclude this WD from further discussion.

\begin{deluxetable*}{rccDDDccc}
\tablecaption{Derived Quantities and Cluster Membership for DA WDs\label{tab.damems}}
\tablewidth{0pt}
\tablehead{ \colhead{ID} & \colhead{$M_{\rm WD}$} & \colhead{$dM_{\rm WD}$} & \twocolhead{$M_g$} & \twocolhead{$dM_g$} & \twocolhead{$\Delta(m-M)_{g}$\tablenotemark{a}} & \colhead{$d(m-M)_g$} & \colhead{PM member\tablenotemark{b,c}} & \colhead{Phot. member\tablenotemark{b}} \\
 & \colhead{$(M_\sun)$} & \colhead{$(M_\sun)$} & \twocolhead{mag} & \twocolhead{mag} & \twocolhead{mag} & mag & & }
\decimals
\startdata
WD3 & 0.696 & 0.032 & 12.213 & 0.084 & -0.118 & 0.090 & Y & Y \\
WD6 & 0.591 & 0.056 & 12.657 & 0.143 & -0.318 & 0.147 & Y & Y \\
WD9 & 0.563 & 0.023 & 11.251 & 0.095 & 0.294 & 0.100 & Y & Y \\
WD10 & 0.795 & 0.025 & 12.240 & 0.068 & -0.624 & 0.071 & Y & B \\
WD14 & 0.627 & 0.025 & 11.464 & 0.071 & -0.123 & 0.075 & Y & Y \\
WD16 & 0.621 & 0.043 & 12.226 & 0.113 & 0.131 & 0.117 & Y & Y \\
WD17 & 0.625 & 0.026 & 11.819 & 0.071 & -0.177 & 0.078 & Y & Y \\
WD19 & 0.657\tablenotemark{d} & 0.017\tablenotemark{d} & 11.305\tablenotemark{d} & 0.052\tablenotemark{d} &  0.116\tablenotemark{d} & 0.061\tablenotemark{d} & Y & Y \\
          & 0.694\tablenotemark{e} & 0.022\tablenotemark{e} & 11.633\tablenotemark{e} & 0.058\tablenotemark{e} &  -0.212\tablenotemark{e} & 0.066\tablenotemark{e} & Y & B \\
WD20 & 0.429 & 0.063 & 12.426 & 0.187 & 0.071 & 0.190 & Y & Y \\
WD22 & 0.390 & 0.029 & 11.419 & 0.112 & 0.229 & 0.117 & Y & Y \\
WD25 & 0.612 & 0.009 & 10.529 & 0.028 & 0.151 & 0.043 & Y & Y \\
WD26 & 0.561 & 0.032 & 12.255 & 0.086 & 0.030 & 0.092 & Y & Y \\
WD29 & 0.759 & 0.031 & 12.294 & 0.082 & -0.182 & 0.088 & Y & Y \\
\hline
WD2 & 0.687 & 0.023 & 11.391 & 0.069 & 0.163 & 0.076 & N & Y \\
WD5 & 0.607 & 0.020 & 11.749 & 0.056 & -0.101 & 0.064 & N & Y \\
WD8 & 0.719 & 0.034 & 10.843 & 0.095 & 0.898 & 0.100 & N & N \\
WD15 & 0.548 & 0.017 & 8.128 & 0.094 & 0.769 & 0.099 & Y & N \\
WD32 & 0.561 & 0.014 & 12.173 & 0.036 & -2.847 & 0.048 & N & N \\
\hline
WD11 & 0.527 & 0.041 & 12.397 & 0.109 & -0.721 & 0.111 & N & \nodata \\
WD12 & 0.759 & 0.059 & 13.871 & 0.143 & -2.648 & 0.145 & N & \nodata \\
WD13 & 0.400 & 0.200 & 12.243 & 0.593 & 1.088 & 0.593 & N & \nodata  \\
WD24 & 0.222 & 0.414 & 12.282 & 0.763 & 1.343 & 0.764 & Y & \nodata \\
WD28 & 0.530 & 0.132 & 12.442 & 0.364 & 1.217 & 0.366 & N & \nodata \\
WD33 & 0.795 & 0.182 & 12.703 & 0.521 & -0.822 & 0.524 & Y & \nodata \\
\enddata
\tablecomments{Objects below the first horizontal line are considered field (non-member) WDs. Objects below the second horizontal line have too large of errors in spectral paramters to constrain photometric membership meaningfully.}
\tablenotetext{a}{Difference of WD and cluster apparent distance moduli}
\tablenotetext{b}{``Y" = yes, ``N" = no, ``B" = candidate binary member}
\tablenotetext{c}{from Bellini et al.\ (2010)}
\tablenotetext{d}{Hot solution (see text)}
\tablenotetext{e}{Cool solution (see text)}
\tablenotetext{f}{From \citet{Kepler2015}}
\end{deluxetable*}

\begin{deluxetable}{rcc}
\tablecaption{Cluster Membership for non-DA WDs\label{tab.dbmems}}
\tablewidth{0pt}
\tablehead{ \colhead{ID} & \colhead{Spectral Type} &  \colhead{PM member\tablenotemark{a}} }
\startdata
WD1 & DB & \nodata\tablenotemark{b} \\
WD4 & DB & N \\
WD7 & DB & N \\
WD18 & DB & Y \\
WD21 & DB & Y \\
WD23 & DB: & N \\
WD27 & DC & Y \\
WD30 & DB & Y \\
WD31 & DC & Y \\
\enddata
\tablenotetext{a}{from Bellini et al.\ (2010)}
\tablenotetext{b}{Outside the region studied by Bellini et al.\ (2010)}
\end{deluxetable}

\subsection{DA WD mass distribution\label{sec.wdmass}}
Stellar evolution theory predicts that, for a simple stellar population, single star evolution should produce WD remnants with masses only dependent on the progenitor star's zero-age main sequence mass.  Therefore, all stars with the same progenitor mass should have the same WD mass.  So far, this assertion has been difficult to test via open cluster WD populations, either because the WDs in a given cluster have significantly different profenitor masses or because the sample size is too small. 

The progenitor mass of a WD can be determined from WD parameters, albeit in a strongly model-dependent way.  This methodology is discussed in \citetalias{PCanton2017} and references therein; we especially point out the discussion of the limitations as discussed by \citet{Salaris2009}.  In brief, the derived \teff and \logg of DA WDs translates directly in to a WD mass and cooling age (time since the cessation of the majority of nuclear burning) via WD evolutionary models.  We subtract this cooling age from the star cluster age to get the nuclear lifetime of the progenitor star.  We then use the PARSEC 1.2S stellar evolutionary models \citep{2015MNRAS.452.1068C} to determine the zero-age main sequence mass of a star with this nuclear lifetime.  As we mention above, the discussion below is not strongly dependent on the exact value of the initial masses for each white dwarf but rather the relative values.

Since M67 is an old cluster, the main sequence turnoff mass is changing relatively slowly, and all of the WDs in our spectroscopic sample should have nearly identical progenitor masses; this idea is clearly illustrated in the initial-final mass relation discussion of \citetalias{PCanton2017}.   Our sample of 13 cluster member DAs with well-determined masses therefore permits us to test whether there is any evidence for additional scatter in the WD masses beyond that due to observational errors.

We start with the sample of cluster member DA WDs given in Table \ref{tab.damems}.  We exclude WD19 from consideration since we are unable to choose unambiguously which atmospheric parameter solution is correct, and we exclude WD10 since it is a potential double degenerate.  This leaves us with a sample of $N=11$ WDs.  We calculate the mean of the WD masses weighted by the mass uncertainties in Table \ref{tab.damems} to be $\bar{M}_{\mathrm DA} = 0.594 \pm 0.011 M_\sun$ (formal error on the mean) with a standard deviation $\sigma_{M_{\mathrm DA}}=0.106 M_\sun$. Three WDs are clear outliers in this distribution: WD20 ($M=0.429\pm 0.066 M_\odot$) and WD22 ($M=0.390\pm 0.034$) are significantly less massive than the mean, and  WD29 is significantly more massive than the mean with $M=0.759\pm 0.039M_\odot$.  We discuss these further below. 

\begin{figure}
\begin{center}
\includegraphics[trim={0 2in 0 1in},clip,width=0.9\columnwidth]{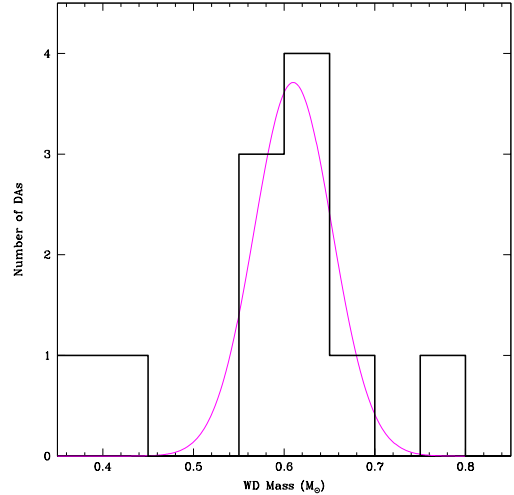}
\end{center}
\caption{The histogram of masses for the 11 well-measured cluster member DA WDs.  The magenta curve is a Gaussian function with a mean value of $0.61 M_\sun$, a standard deviation of $0.043 M_\sun$, and normalized to a total of 8 WDs.  The three outliers are WD20 and WD22 (low-mass end) and WD29 (high-mass end).  When these three outliers are excluded, the measured standard deviation is very close to our calculated WD mass errors, consistent with the hypothesis that all WDs have the same mass to better than a few percent.\label{fig.mass_hist}}
\end{figure}

If we remove the three outlier WDs from the sample (now $N=8$), we find a weighted mean mass $\bar{M}_{\mathrm DA} = 0.610 \pm 0.012 M_\sun$ (formal error on the mean) with a standard deviation $\sigma_{M_{\mathrm DA}}=0.043 M_\sun$; the distribution is consistent with a Gaussian with the same standard deviation (see Figure \ref{fig.mass_hist}).  The mean mass is $\approx 0.08M_\sun$ higher than that predicted by the linear semi-empirical initial-final mass relation of \citet{2009ApJ...693..355W} of $0.53M_\sun$ for a $1.5M_\odot$ progenitor mass; this is discussed further in \citetalias{PCanton2017}.  

The standard deviation of $0.043M_\sun$ is very similar to the average of the WD mass measurement uncertainties ($0.039 M_\sun$), suggesting any additional source of scatter in the WD mass distribution beyond observational uncertainties must be $\lesssim 0.02 M_\sun$, assuming the intrinsic scatter and observational errors can be added in quadrature.   In other words, single stars in M67 near the current main sequence turnoff mass result in WD remnants with an intrinsic scatter of less than 3\% in the WD mass  -- if we have estimated our observational errors correctly.  If our mass errors are significantly overestimated, then the intrinsic scatter could be as large as the observed scatter of $\approx 0.04 M_\sun$, or $\approx 7\%$.  

\begin{figure*}
\begin{center}
\begin{minipage}{0.49\textwidth}
\includegraphics[trim={0 2in 0 1in},clip,width=0.99\textwidth]{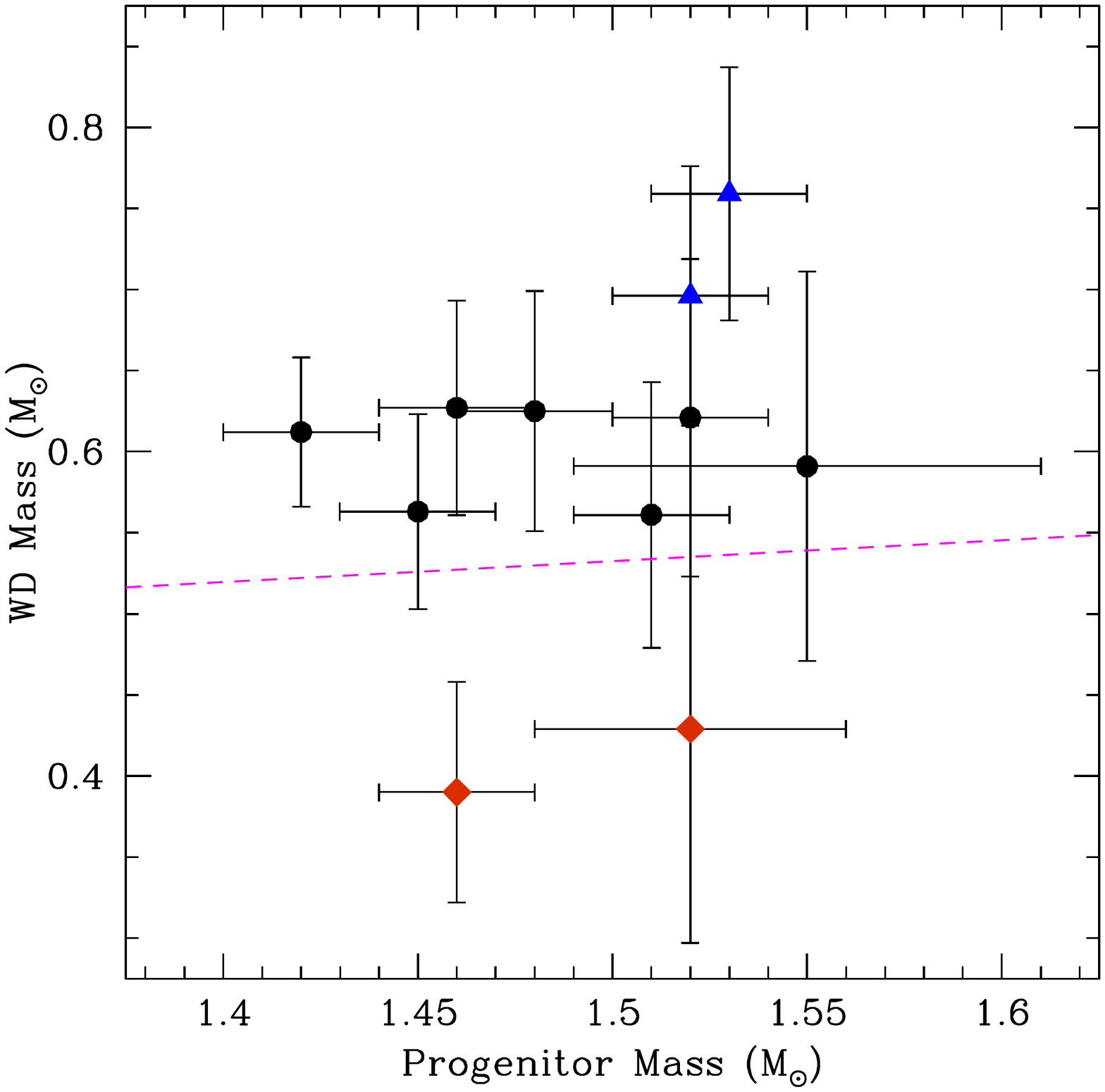}
\end{minipage}\hfill
\begin{minipage}{0.49\textwidth}
\includegraphics[trim={0 2in 0 1in},clip,width=0.99\textwidth]{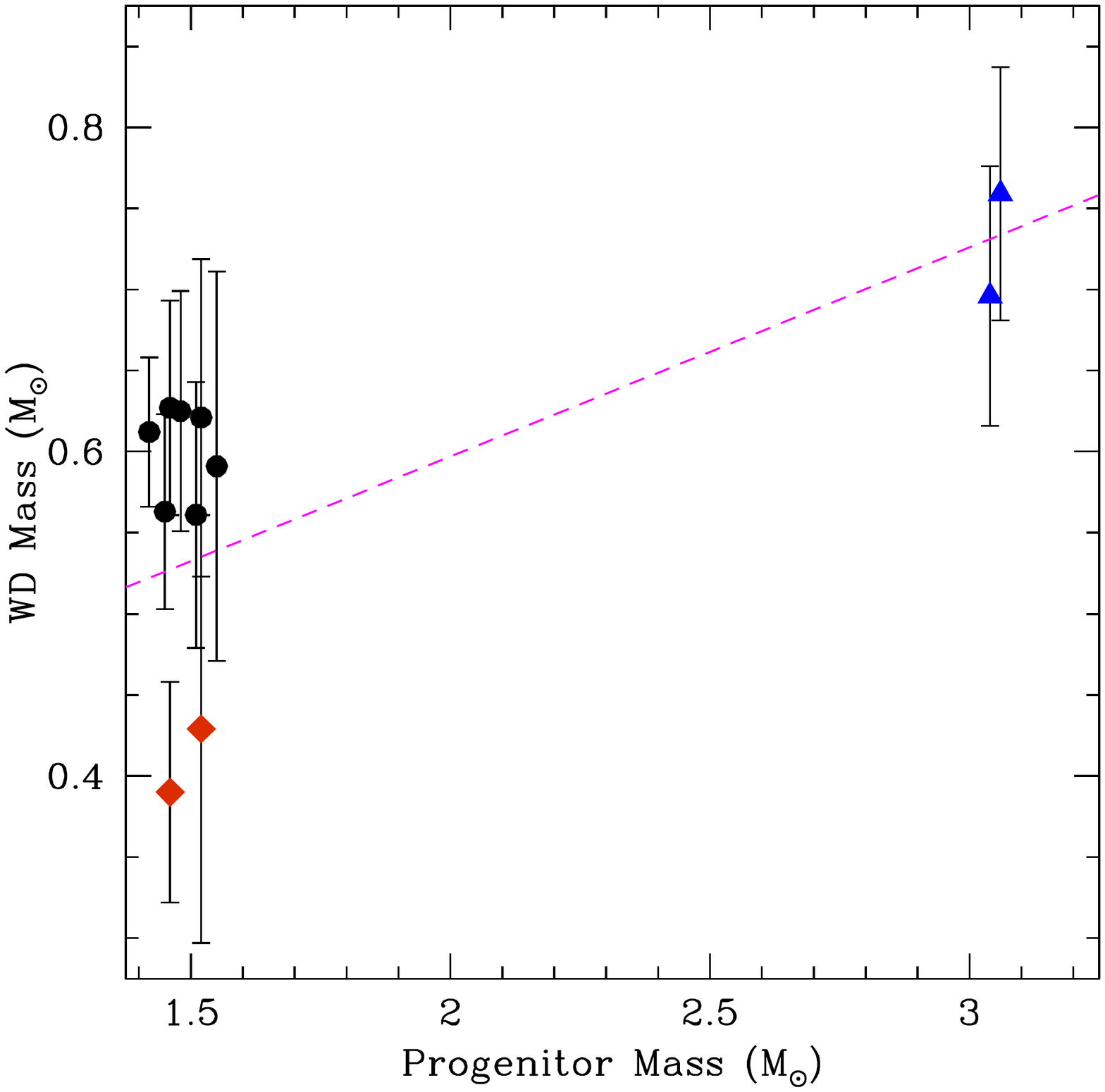}
\end{minipage}
\end{center}
\caption{The initial-final mass relation for M67 DA WDs.  The left panel assumes the initial masses as determined in \citetalias{PCanton2017}; the right panel assumes twice the calculated initial mass for the two blue straggler candidates.  Points indicate the location of the 11 well-measured cluster member DAs; plotted error bars are $2\sigma$.  Brown diamonds indicate the two candidate He-core WDs WD20 and WD22; blue trangles are the two candidate blue straggler remnants WD29 and WD3.  The magenta line is the linear fit to the semi-empirical initial-final mass relation of \citet{2009ApJ...693..355W}.  The WD masses have no significant correlation with their progenitor star masses over this tiny initial mass range, but the initial-final mass relation slope is also quite flat over the same range of initial masses.  The functional fit also clearly underpredicts the DA WD masses, as discussed in \citetalias{PCanton2017}; this holds true for other semi-empirical IFMR fits such as \citet{2016ApJ...818...84C}. The blue straggler progeny agree with the plotted initial-final mass relation in the right-hand panel, suggesting that after mass transfer these blue stragglers evolved similarly to a single star of the same mass. \label{fig.ifmr}}
\end{figure*}

Overall, one should expect that cooler WDs in a star cluster should have a larger mass than hotter WDs, since the cooler WDs came from progenitor stars with shorter nuclear lifetimes, implying a higher progenitor mass.  Indeed, this effect is seen in younger open clusters such as NGC 2168 and NGC 2099 \citep{2009ApJ...693..355W,2005ApJ...618L.123K,2015ApJ...807...90C}.  Given the advanced age of M67 and the comparatively short cooling time of our spectroscopic sample, the derived progenitor masses for our WDs span less than $0.1M_\sun$, as seen in \citetalias{PCanton2017}.  Based on linear approximations to the semiempirical initial-final mass relation \citep[e.g.,][]{2009ApJ...693..355W,2016ApJ...818...84C}, one would expect only a change in WD masses of $\approx 0.015 M_\odot$ over the M67 sample initial mass range, significantly smaller than our observed scatter.  Indeed, no significant correlation is found between the M67 WD masses and their initial mass -- see Figure \ref{fig.ifmr}.

\subsection{Two Candidate He-core WDs} 
As mentioned above, WD20 and WD22 are significantly less massive than the cluster mean.  In fact, both WDs are less massive than the canonical mass required for He-core ignition of $\approx 0.45M_\odot$ \citep[e.g.,]{1994ApJ...426..612S,2001PASP..113..409F}, though these are only $\approx 2\sigma$ deviations.   This would imply that a fraction $\approx 0.18$ of M67 WDs are He-core WDs, with a 90\% upper confidence limit of $\approx 0.4$ assuming a binomial distribution.

Candidate He-core WDs have also been identified in the older, very metal rich open cluster NGC 6791 by \citet{Kalirai2007}, who present evidence of a He-core WD fraction of greater 50\%.  They propose that mass loss during the first ascent of the red giant branch may be responsible for the He-core WDs in that cluster, though \citet{2012MNRAS.419.2077M} find no evidence of such extreme mass loss by red giants in NGC 6791.  \citet{Kalirai2007} also propose that the high He-core WD fraction is likely related to the known extreme horizontal branch (EHB) stars known to be in the cluster.

Our measurement of the M67 He-core WD fraction is inconsistent with that observed in NGC 6791, and recent K2 asteroseismology results on the masses of giants in M67 likewise find no evidence for extreme mass loss on the red giant branch \citep{2016ApJ...832..133S}.  M67 has no known EHB stars \citep{1994AJ....107.1408L}.  Assuming that WD20 and/or WD22 are bona-fide He-core WDs, it seems likely that either the He-core WDs in M67 form by a different mechanism than those in NGC 6791, or that the formation mechanism in both clusters is different than that proposed by \citet{Kalirai2007}.

Another mechanism for the formation of He-core WDs is through binary interactions.  Our spectra are too low of resolution to resolve radial velocity variations, but we can compare synthetic and observed photometry for inconsistencies with an appropriatelly reddened single WD model.  In particular, we compare the color indices $\Delta(u-g)=(u-g)_{\mathrm WD} - (u-g)_{\mathrm model}$ and $\Delta(g-r)=(g-r)_{\mathrm WD} - (g-r)_{\mathrm model}$.  WD20 has color indices fully consistent with the single star model: $\Delta(u-g)=0.018\pm 0.042$ and $\Delta(g-r)=0.0040\pm 0.047$.  Therefore, any companion to WD20 must be extremely faint in our bandpasses.

WD22, on the other hand, does exhibit significantly different colors than predicted by synthetic photometry: $\Delta(u-g)=-0.145\pm 0.040$ and $\Delta(g-r)=+0.328\pm 0.052$.  The significantly redder $g-r$ color could imply a faint, cool companion.  No evidence of an M dwarf companion is obvious in the spectrum (see Figure \ref{fig.daspec}), but the spectrum is relatively noisy and could hide a companion similar to that observed in the foreground WD+dM system WD32, which has a similar color excess. 
It is thus possible that the candidate He-core WD22 has an M dwarf companion.  

\subsection{WD(s) from Blue Stragglers} 
The other WD with a significantly discrepant mass is WD29, with $M_{\rm WD}=0.759\pm 0.039 M_\sun$.  We propose that this WD is the remnant of a blue straggler.  M67 has a significant blue straggler population, some of which have measured masses.  For example, S1237 is an evolved blue straggler with an asteroseismic mass of $2.9\pm 0.2 M_\sun$ \citep{2016ApJ...832L..13L}, and S1082 is a triple star system containing at least two blue stragglers, including S1082 Aa with a mass of $2.52\pm 0.38 M_\sun$ \citep{2003AJ....125..810S}.  

For the sake of argument, let us make the assumption that a blue straggler will form an ordinary C/O WD identical to the WD remnant of a single star with the blue straggler's mass.  The WD resulting from a $2.9M_\sun$ star such as S1237 would have a mass of $0.71 M_\sun$ based on the initial-final mass relations of \citet{2009ApJ...693..355W} and \citet{2016ApJ...818...84C}.  This is consistent with the mass of WD29.   Interestingly, this is also consistent with the mass of WD3 ($M_{\mathrm WD}=0.693\pm 0.040 M_\sun$), which otherwise exhibits moderate tension ($\approx 2\sigma$) with the mean mass of the cluster WDs.  In the right-hand panel of Figure \ref{fig.ifmr}, we show the initial-final mass plane for M67 WDs if we assume that the progenitor masses of WD3 and WD29 were twice their value as calculated in \citetalias{PCanton2017}, similar to if they arose from blue stragglers; both WD's masses are fully consistent with the initial-final mass relation at this doubled progenitor mass.

It therefore is plausible that at least one and perhaps two of the DA WDs in M67 arose from blue stragglers, and that the WD mass resulting from blue straggler evolution is not significantly different from what might be expected from the evolution of a single star with a mass equal to that of the blue straggler.  

\subsection{DA:non-DA ratio}
We calculate the ratio of DA to non-DA spectral types as follows.  For the DA spectral types, we consider all proper motion members consistent with being single WDs in the cluster; i.e., we exclude the potential binary WD10 and the photometrically inconsistent WD15.  We also exclude WD24 and WD33, proper motion members whose spectra are too noisy to permit photometric membership determinations.  For this case, we include WD19, the DA for which we could not select between hot and cold temperature solutions, since its spectral type is not in doubt.  Our sample thus contains 12 DAs with $g\leq 22.2$.  For non-DA spectral types, we include all proper motion members with $g\leq 22.2$; i.e., in the same apparent magnitude range as the DAs.  Our sample thus consists of four non-DAs: WD18, WD21, WD30, and WD31.  
To first order, this gives us a DA:non-DA ratio of $4:1$, nearly identical to that in the field, though with significant uncertainty ($\sim 50\%$) once small number statistics are considered.  Given the high spectroscopic completeness for our post facto selected candidates, any effects due to incompleteness are far smaller than those due to small number statistics.

\deleted{As described above, the target selection included all stellar objects  over a fairly wide color range that includes both DA and non-DA spectral types, with some weighting as a function of apparent magnitude.   WDs with H-dominated and He-dominated atmospheres cool at different rates over different magnitude ranges.  Taking $g=22.2\,(M_g\approx 12.4)$ as our magnitude limit and using the $0.6 M_\sun$ evolutionary models of \citet{2001PASP..113..409F}, we find that the cooling time of a DA at our magnitude limit is 698 Myr, while the cooling time for a DB of this magnitude is 696 Myr, nearly identical.   We note that this agreement is a coincidence, and a detailed correction would need to convolve cooling ages and completeness as a function of magnitude.  However, given our small number statistics, such a calculation would not significantly change the results, so we do not attempt it here.}

We note in \citet{2006ApJ...643L.127W} that an appropriate accounting of non-DA WDs in a given open cluster requires consideration of the so-called ``DB gap,'' a noted change in the DA:non-DA ratio for WDs with $\teff\approx 30,000$  to $45,000$ K.  The DB cooling models of \citet{2001PASP..113..409F} show that a $0.6M_\sun$ WD will cool below 30,000 K in $\approx 14$ Myr, or roughly 2\% of the cooling time of WDs at our limiting magnitude.  Therefore, any DB gap correction will be on the order of this level, and insignificant compared to our statistical errors.

Therefore, we find no evidence that the DA to non-DA ratio differs in M67 as compared to the ratio in the field.  This suggests that any proposed mechanism for suppressing non-DA formation in the cluster environment \citep[e.g., as suggested in][]{2009ApJ...705..398D} would need to explain this exception.

\section{Summary}

In this paper, we present the results of a spectroscopic study of the WD population in the field of the open cluster M67.  A combination of proper motion and photometric distance determinations allows us to study an uncontaminated  sample of cluster member WDs, including 13 DA (hydrogen-dominated atmosphere) WDs and four non-DA (helium-dominated atmosphere) WDs.  Our spectroscopy is highly complete for non-binary cluster WDs; $\sim 90\%$ of photometrically selected candidates have spectroscopy, and 75\% of the DAs in our sample have spectra of sufficient quality to permit photometric membership determination.  While the photometric completeness is not quantified, it is likely to be high and relatively unbiased for $g\leq 22.2\,(M_g\approx 12.4)$.  Because M67 is an old open cluster of solar metallicity, this sample provides unique insight into the end stages of stellar evolution for solar-type stars.  

In particular, we find the following:
\begin{itemize}
\item The mass distribution of DA WDs suggests that any scatter in integrated mass loss over the lifetime of a single, $\approx 1.5 M_\sun$ star is small -- less than 7\% of the WD mass, \emph{if} our interpretation that the three significant outliers to this distribution are the result of binary evolution is correct. \deleted{, and potentially $\lesssim 3\%$ of the resulting WD mass if our errors are properly characterized and of a Gaussian distribution.}  We find the average mass of the DA WDs to be $0.610\pm 0.012 M_\sun$ with a standard deviation of $0.043 M_\sun$, comparable with our typical observational uncertainty in mass determination of $0.039 M_\sun$.

\item We identify two cluster member DA WDs that are consistent with being He-core WDs, for a He-core WD fraction of $\sim 20\%$, in contrast to others' findings for the WDs in the super metal rich cluster NGC 6791.  Other researchers find no observational evidence for strong mass loss on the M67 red giant branch, and M67 contains no extreme horizontal branch stars.  If high metallicity can cause extreme mass loss on the red giant branch, the enhanced mass loss mechanism is not efficient at or below solar metalicity.

\item One or two M67 DA WDs have masses significantly higher than the cluster DA WD mean mass.  These WDs are potential progeny of blue straggler stars.  The WD masses are consistent with the expectation from the initial-final mass relation if their progenitor stars were twice as massive as expected from single-star evolution.

\item The ratio of hydrogen- to helium-dominated atmosphere WDs (DA to non-DA ratio) is $\sim 4:1$, consistent with that measured for field WDs.  This argues against the proposal that a mechanism in clusters inhibits formation of non-DA WDs, at least for solar metallicity, low mass stars.

\item We identify one DA WD that is a proper motion member of M67 but overluminous, indicating it is a candidate unresolved double degenerate.

\end{itemize}

\acknowledgments
The authors wish to recognize and acknowledge the very significant cultural role and reverence that the summit of Maunakea has always had within the indigenous Hawaiian community.  We are most fortunate to have the opportunity to conduct observations from this mountain. The authors thank P.~Bergeron for discussions related to synthetic photometry, J.~Kalirai for accompanying us on our clouded-out 2006 observing run, and L.~Bedin for logistical support.  The authors appreciate discussions and encouragement related to this work especially from J.~Liebert, R.~Falcon, K.~Montgomery, and M.~Wood.  K.A.W. acknowledges the support of NSF award AST-0602288 during the data collection of this project, as well as the support of a Cottrell College Science Award from the Research Corporation for Science Advancement.  M.K. acknowledges support of NSF award AST-1312678.  This research has made use of NASA's Astrophysics Data System and of the SIMBAD database, operated at CDS, Strasbourg, France.  The authors are grateful for the helpful suggestions of the referee of this manuscript. Finally, the authors wish to acknowledge the great and positive impact that Enrique Garc'a-Berro had on the field of white dwarf astrophysics and the people around him.  We mourn his loss.

\facility{Keck:I (LRIS-B)},\facility{MMT (Megacam)}

\software{IRAF \citep{1986SPIE..627..733T,1993ASPC...52..173T}, SExtractor \citep{1996A&AS..117..393B}, SWarp \citep{2002ASPC..281..228B}, DAOPHOT II \citep{1987PASP...99..191S}, L.A.Cosmic \citep{2001PASP..113.1420V}}


\begin{thebibliography}{}
\expandafter\ifx\csname natexlab\endcsname\relax\def\natexlab#1{#1}\fi
\providecommand{\url}[1]{\href{#1}{#1}}

\bibitem[{{Alam} {et~al.}(2015){Alam}, {Albareti}, {Allende Prieto}, {Anders},
  {Anderson}, {Anderton}, {Andrews}, {Armengaud}, {Aubourg}, {Bailey}, \&
  et~al.}]{2015ApJS..219...12A}
{Alam}, S., {Albareti}, F.~D., {Allende Prieto}, C., {et~al.} 2015, \apjs, 219,
  12

\bibitem[{Babusiaux {et~al.}(2018)Babusiaux, van Leeuwen, Barstow, Jordi,
  Vallenari, Bossini, Bressan, \& and}]{Babusiaux2018}
Babusiaux, C., van Leeuwen, F., Barstow, M., {et~al.} 2018, Astronomy {\&}
  Astrophysics, doi:10.1051/0004-6361/201832843, in press

\bibitem[{{B{\'e}dard} {et~al.}(2017){B{\'e}dard}, {Bergeron}, \&
  {Fontaine}}]{2017arXiv170902324B}
{B{\'e}dard}, A., {Bergeron}, P., \& {Fontaine}, G. 2017, \apj, 848, 11

\bibitem[{{Bedin} {et~al.}(2008{\natexlab{a}}){Bedin}, {King}, {Anderson},
  {Piotto}, {Salaris}, {Cassisi}, \& {Serenelli}}]{2008ApJ...678.1279B}
{Bedin}, L.~R., {King}, I.~R., {Anderson}, J., {et~al.} 2008{\natexlab{a}},
  \apj, 678, 1279

\bibitem[{{Bedin} {et~al.}(2008{\natexlab{b}}){Bedin}, {Salaris}, {Piotto},
  {Cassisi}, {Milone}, {Anderson}, \& {King}}]{2008ApJ...679L..29B}
{Bedin}, L.~R., {Salaris}, M., {Piotto}, G., {et~al.} 2008{\natexlab{b}},
  \apjl, 679, L29

\bibitem[{{Bedin} {et~al.}(2005){Bedin}, {Salaris}, {Piotto}, {King},
  {Anderson}, {Cassisi}, \& {Momany}}]{2005ApJ...624L..45B}
---. 2005, \apjl, 624, L45

\bibitem[{{Bellini} {et~al.}(2010){Bellini}, {Bedin}, {Piotto}, {Salaris},
  {Anderson}, {Brocato}, {Ragazzoni}, {Ortolani}, {Bonanos}, {Platais},
  {Gilliland}, {Raimondo}, {Bragaglia}, {Tosi}, {Gallozzi}, {Testa},
  {Kochanek}, {Giallongo}, {Baruffolo}, {Farinato}, {Diolaiti}, {Speziali},
  {Carraro}, \& {Yadav}}]{2010A&A...513A..50B}
{Bellini}, A., {Bedin}, L.~R., {Piotto}, G., {et~al.} 2010, \aap, 513, A50

\bibitem[{{Bergeron} {et~al.}(1992){Bergeron}, {Saffer}, \&
  {Liebert}}]{Bergeron1992}
{Bergeron}, P., {Saffer}, R.~A., \& {Liebert}, J. 1992, \apj, 394, 228

\bibitem[{Bergeron {et~al.}(1995)Bergeron, Wesemael, Lamontagne, Fontaine,
  Saffer, \& Allard}]{1995ApJ...449..258B}
Bergeron, P., Wesemael, F., Lamontagne, R., {et~al.} 1995, \apj, 449, 258

\bibitem[{{Bergeron} {et~al.}(2011){Bergeron}, {Wesemael}, {Dufour},
  {Beauchamp}, {Hunter}, {Saffer}, {Gianninas}, {Ruiz}, {Limoges}, {Dufour},
  {Fontaine}, \& {Liebert}}]{2011ApJ...737...28B}
{Bergeron}, P., {Wesemael}, F., {Dufour}, P., {et~al.} 2011, \apj, 737, 28

\bibitem[{{Bertin} \& {Arnouts}(1996)}]{1996A&AS..117..393B}
{Bertin}, E., \& {Arnouts}, S. 1996, \aaps, 117, 393

\bibitem[{{Bertin} {et~al.}(2002){Bertin}, {Mellier}, {Radovich}, {Missonnier},
  {Didelon}, \& {Morin}}]{2002ASPC..281..228B}
{Bertin}, E., {Mellier}, Y., {Radovich}, M., {et~al.} 2002, in Astronomical
  Society of the Pacific Conference Series, Vol. 281, Astronomical Data
  Analysis Software and Systems XI, ed. D.~A. {Bohlender}, D.~{Durand}, \&
  T.~H. {Handley}, 228

\bibitem[{Bonatto {et~al.}(2015)Bonatto, Campos, Kepler, \&
  Bica}]{2015MNRAS.450.2500B}
Bonatto, C., Campos, F., Kepler, S.~O., \& Bica, E. 2015, \mnras, 450, 2500

\bibitem[{{Brown} {et~al.}(2010){Brown}, {Kilic}, {Allende Prieto}, \&
  {Kenyon}}]{2010ApJ...723.1072B}
{Brown}, W.~R., {Kilic}, M., {Allende Prieto}, C., \& {Kenyon}, S.~J. 2010,
  \apj, 723, 1072

\bibitem[{{Canton} {et~al.}(2018){Canton}, {Williams}, {Kilic}, \&
  {Bolte}}]{PCanton2017}
{Canton}, P., {Williams}, K., {Kilic}, M., \& {Bolte}, M. 2018, submitted to
  MNRAS

\bibitem[{{Chen} \& {Han}(2008)}]{2008MNRAS.387.1416C}
{Chen}, X., \& {Han}, Z. 2008, \mnras, 387, 1416

\bibitem[{{Chen} {et~al.}(2015){Chen}, {Bressan}, {Girardi}, {Marigo}, {Kong},
  \& {Lanza}}]{2015MNRAS.452.1068C}
{Chen}, Y., {Bressan}, A., {Girardi}, L., {et~al.} 2015, \mnras, 452, 1068

\bibitem[{Chen {et~al.}(2014)Chen, Girardi, Bressan, Marigo, Barbieri, \&
  Kong}]{2014MNRAS.444.2525C}
Chen, Y., Girardi, L., Bressan, A., {et~al.} 2014, \mnras, 444, 2525

\bibitem[{{Claver} {et~al.}(2001){Claver}, {Liebert}, {Bergeron}, \&
  {Koester}}]{2001ApJ...563..987C}
{Claver}, C.~F., {Liebert}, J., {Bergeron}, P., \& {Koester}, D. 2001, \apj,
  563, 987

\bibitem[{{Cummings} {et~al.}(2015){Cummings}, {Kalirai}, {Tremblay}, \&
  {Ramirez-Ruiz}}]{2015ApJ...807...90C}
{Cummings}, J.~D., {Kalirai}, J.~S., {Tremblay}, P.-E., \& {Ramirez-Ruiz}, E.
  2015, \apj, 807, 90

\bibitem[{Cummings {et~al.}(2016)Cummings, Kalirai, Tremblay, \&
  Ramirez-Ruiz}]{2016ApJ...818...84C}
Cummings, J.~D., Kalirai, J.~S., Tremblay, P.-E., \& Ramirez-Ruiz, E. 2016,
  \apj, 818, 84

\bibitem[{{Davis} {et~al.}(2009){Davis}, {Richer}, {Rich}, {Reitzel}, \&
  {Kalirai}}]{2009ApJ...705..398D}
{Davis}, D.~S., {Richer}, H.~B., {Rich}, R.~M., {Reitzel}, D.~R., \& {Kalirai},
  J.~S. 2009, \apj, 705, 398

\bibitem[{{Dobbie} {et~al.}(2012){Dobbie}, {Day-Jones}, {Williams}, {Casewell},
  {Burleigh}, {Lodieu}, {Parker}, \& {Baxter}}]{Dobbie2012}
{Dobbie}, P.~D., {Day-Jones}, A., {Williams}, K.~A., {et~al.} 2012, \mnras,
  423, 2815

\bibitem[{{Dobbie} {et~al.}(2009){Dobbie}, {Napiwotzki}, {Burleigh},
  {Williams}, {Sharp}, {Barstow}, {Casewell}, \& {Hubeny}}]{Dobbie2009}
{Dobbie}, P.~D., {Napiwotzki}, R., {Burleigh}, M.~R., {et~al.} 2009, \mnras,
  395, 2248

\bibitem[{{Fan} {et~al.}(1996){Fan}, {Burstein}, {Chen}, {Zhu}, {Jiang}, {Wu},
  {Yan}, {Zheng}, {Zhou}, {Fang}, {Chen}, {Deng}, {Chu}, {Hester}, {Windhorst},
  {Li}, {Lu}, {Sun}, {Chen}, {Tsay}, {Chiueh}, {Chou}, {Ko}, {Lin}, {Guo}, \&
  {Byun}}]{1996AJ....112..628F}
{Fan}, X., {Burstein}, D., {Chen}, J.-S., {et~al.} 1996, \aj, 112, 628

\bibitem[{{Fleming} {et~al.}(1997){Fleming}, {Liebert}, {Bergeron}, \&
  {Beauchamp}}]{1997ASSL..214...91F}
{Fleming}, T.~A., {Liebert}, J., {Bergeron}, P., \& {Beauchamp}, A. 1997, in
  Astrophysics and Space Science Library, Vol. 214, White dwarfs, ed.
  J.~{Isern}, M.~{Hernanz}, \& E.~{Garcia-Berro}, 91

\bibitem[{{Fontaine} {et~al.}(2001){Fontaine}, {Brassard}, \&
  {Bergeron}}]{2001PASP..113..409F}
{Fontaine}, G., {Brassard}, P., \& {Bergeron}, P. 2001, \pasp, 113, 409

\bibitem[{{Fontaine} \& {Wesemael}(1987)}]{1987fbs..conf..319F}
{Fontaine}, G., \& {Wesemael}, F. 1987, in IAU Colloq. 95: Second Conference on
  Faint Blue Stars, ed. A.~G.~D. {Philip}, D.~S. {Hayes}, \& J.~W. {Liebert},
  319--326

\bibitem[{{Garc{\'{\i}}a-Berro} {et~al.}(2010){Garc{\'{\i}}a-Berro}, {Torres},
  {Althaus}, {Renedo}, {Lor{\'e}n-Aguilar}, {C{\'o}rsico}, {Rohrmann},
  {Salaris}, \& {Isern}}]{2010Natur.465..194G}
{Garc{\'{\i}}a-Berro}, E., {Torres}, S., {Althaus}, L.~G., {et~al.} 2010, \nat,
  465, 194

\bibitem[{{Geller} \& {Mathieu}(2011)}]{2011Natur.478..356G}
{Geller}, A.~M., \& {Mathieu}, R.~D. 2011, \nat, 478, 356

\bibitem[{{Gianninas} {et~al.}(2011){Gianninas}, {Bergeron}, \&
  {Ruiz}}]{Gianninas2011}
{Gianninas}, A., {Bergeron}, P., \& {Ruiz}, M.~T. 2011, \apj, 743, 138

\bibitem[{{Gianninas} {et~al.}(2015){Gianninas}, {Kilic}, {Brown}, {Canton}, \&
  {Kenyon}}]{2015ApJ...812..167G}
{Gianninas}, A., {Kilic}, M., {Brown}, W.~R., {Canton}, P., \& {Kenyon}, S.~J.
  2015, \apj, 812, 167

\bibitem[{{Gilliland} {et~al.}(1991){Gilliland}, {Brown}, {Duncan}, {Suntzeff},
  {Lockwood}, {Thompson}, {Schild}, {Jeffrey}, \&
  {Penprase}}]{1991AJ....101..541G}
{Gilliland}, R.~L., {Brown}, T.~M., {Duncan}, D.~K., {et~al.} 1991, \aj, 101,
  541

\bibitem[{{Gosnell} {et~al.}(2014){Gosnell}, {Mathieu}, {Geller}, {Sills},
  {Leigh}, \& {Knigge}}]{2014ApJ...783L...8G}
{Gosnell}, N.~M., {Mathieu}, R.~D., {Geller}, A.~M., {et~al.} 2014, \apjl, 783,
  L8

\bibitem[{{Gratton} {et~al.}(2006){Gratton}, {Bragaglia}, {Carretta}, \&
  {Tosi}}]{2006ApJ...642..462G}
{Gratton}, R., {Bragaglia}, A., {Carretta}, E., \& {Tosi}, M. 2006, \apj, 642,
  462

\bibitem[{{Hansen}(2005)}]{2005ApJ...635..522H}
{Hansen}, B.~M.~S. 2005, \apj, 635, 522

\bibitem[{{Holberg} \& {Bergeron}(2006)}]{2006AJ....132.1221H}
{Holberg}, J.~B., \& {Bergeron}, P. 2006, \aj, 132, 1221

\bibitem[{{Hurley} {et~al.}(2005){Hurley}, {Pols}, {Aarseth}, \&
  {Tout}}]{2005MNRAS.363..293H}
{Hurley}, J.~R., {Pols}, O.~R., {Aarseth}, S.~J., \& {Tout}, C.~A. 2005,
  \mnras, 363, 293

\bibitem[{{Hurley} \& {Shara}(2002)}]{2002ApJ...570..184H}
{Hurley}, J.~R., \& {Shara}, M.~M. 2002, \apj, 570, 184

\bibitem[{{Iben} {et~al.}(1983){Iben}, {Kaler}, {Truran}, \&
  {Renzini}}]{1983ApJ...264..605I}
{Iben}, Jr., I., {Kaler}, J.~B., {Truran}, J.~W., \& {Renzini}, A. 1983, \apj,
  264, 605

\bibitem[{{Iben} \& {Livio}(1993)}]{1993PASP..105.1373I}
{Iben}, Jr., I., \& {Livio}, M. 1993, \pasp, 105, 1373

\bibitem[{{Kalirai} {et~al.}(2007){Kalirai}, {Bergeron}, {Hansen}, {Kelson},
  {Reitzel}, {Rich}, \& {Richer}}]{Kalirai2007}
{Kalirai}, J.~S., {Bergeron}, P., {Hansen}, B.~M.~S., {et~al.} 2007, \apj, 671,
  748

\bibitem[{{Kalirai} {et~al.}(2005{\natexlab{a}}){Kalirai}, {Richer}, {Hansen},
  {Reitzel}, \& {Rich}}]{2005ApJ...618L.129K}
{Kalirai}, J.~S., {Richer}, H.~B., {Hansen}, B.~M.~S., {Reitzel}, D., \&
  {Rich}, R.~M. 2005{\natexlab{a}}, \apjl, 618, L129

\bibitem[{{Kalirai} {et~al.}(2005{\natexlab{b}}){Kalirai}, {Richer}, {Reitzel},
  {Hansen}, {Rich}, {Fahlman}, {Gibson}, \& {von Hippel}}]{2005ApJ...618L.123K}
{Kalirai}, J.~S., {Richer}, H.~B., {Reitzel}, D., {et~al.} 2005{\natexlab{b}},
  \apjl, 618, L123

\bibitem[{{Kalirai} {et~al.}(2009){Kalirai}, {Saul Davis}, {Richer},
  {Bergeron}, {Catelan}, {Hansen}, \& {Rich}}]{2009ApJ...705..408K}
{Kalirai}, J.~S., {Saul Davis}, D., {Richer}, H.~B., {et~al.} 2009, \apj, 705,
  408

\bibitem[{{Kalirai} {et~al.}(2001){Kalirai}, {Richer}, {Fahlman}, {Cuillandre},
  {Ventura}, {D'Antona}, {Bertin}, {Marconi}, \&
  {Durrell}}]{2001AJ....122..257K}
{Kalirai}, J.~S., {Richer}, H.~B., {Fahlman}, G.~G., {et~al.} 2001, \aj, 122,
  257

\bibitem[{Kepler {et~al.}(2015)Kepler, Pelisoli, Koester, Ourique, Kleinman,
  Romero, Nitta, Eisenstein, Costa, K{\"u}lebi, Jordan, Dufour, Giommi, \&
  Rebassa-Mansergas}]{Kepler2015}
Kepler, S.~O., Pelisoli, I., Koester, D., {et~al.} 2015, \mnras, 446, 4078

\bibitem[{Kilic {et~al.}(2007)Kilic, Stanek, \&
  Pinsonneault}]{2007ApJ...671..761K}
Kilic, M., Stanek, K.~Z., \& Pinsonneault, M.~H. 2007, \apj, 671, 761

\bibitem[{{Koester}(2010)}]{Koester2010}
{Koester}, D. 2010, \memsai, 81, 921

\bibitem[{{Koester} \& {Kepler}(2015)}]{2015A&A...583A..86K}
{Koester}, D., \& {Kepler}, S.~O. 2015, \aap, 583, A86

\bibitem[{{Kowalski} \& {Saumon}(2006)}]{2006ApJ...651L.137K}
{Kowalski}, P.~M., \& {Saumon}, D. 2006, \apjl, 651, L137

\bibitem[{{Landsman} {et~al.}(1997){Landsman}, {Aparicio}, {Bergeron}, {Di
  Stefano}, \& {Stecher}}]{1997ApJ...481L..93L}
{Landsman}, W., {Aparicio}, J., {Bergeron}, P., {Di Stefano}, R., \& {Stecher},
  T.~P. 1997, \apjl, 481, L93

\bibitem[{{Landsman} {et~al.}(1998){Landsman}, {Bohlin}, {Neff}, {O'Connell},
  {Roberts}, {Smith}, \& {Stecher}}]{1998AJ....116..789L}
{Landsman}, W., {Bohlin}, R.~C., {Neff}, S.~G., {et~al.} 1998, \aj, 116, 789

\bibitem[{{Leiner} {et~al.}(2016){Leiner}, {Mathieu}, {Stello}, {Vanderburg},
  \& {Sandquist}}]{2016ApJ...832L..13L}
{Leiner}, E., {Mathieu}, R.~D., {Stello}, D., {Vanderburg}, A., \& {Sandquist},
  E. 2016, \apjl, 832, L13

\bibitem[{{Liebert} {et~al.}(2005){Liebert}, {Bergeron}, \&
  {Holberg}}]{2005ApJS..156...47L}
{Liebert}, J., {Bergeron}, P., \& {Holberg}, J.~B. 2005, \apjs, 156, 47

\bibitem[{{Liebert} {et~al.}(1994){Liebert}, {Saffer}, \&
  {Green}}]{1994AJ....107.1408L}
{Liebert}, J., {Saffer}, R.~A., \& {Green}, E.~M. 1994, \aj, 107, 1408

\bibitem[{{Luyten}(1963)}]{1963LB....C32....1L}
{Luyten}, W.~J. 1963, in The Observatory, Univ. Minnesota, Minneapolis, 1953,
  Vol. 32, p. 1 (1963), Vol.~32, 1

\bibitem[{{Marsh} {et~al.}(1995){Marsh}, {Dhillon}, \&
  {Duck}}]{1995MNRAS.275..828M}
{Marsh}, T.~R., {Dhillon}, V.~S., \& {Duck}, S.~R. 1995, \mnras, 275, 828

\bibitem[{{Mathieu} \& {Latham}(1986)}]{1986AJ.....92.1364M}
{Mathieu}, R.~D., \& {Latham}, D.~W. 1986, \aj, 92, 1364

\bibitem[{{McCarthy} {et~al.}(1998){McCarthy}, {Cohen}, {Butcher}, {Cromer},
  {Croner}, {Douglas}, {Goeden}, {Grewal}, {Lu}, {Petrie}, {Weng}, {Weber},
  {Koch}, \& {Rodgers}}]{1998SPIE.3355...81M}
{McCarthy}, J.~K., {Cohen}, J.~G., {Butcher}, B., {et~al.} 1998, in \procspie,
  Vol. 3355, Optical Astronomical Instrumentation, ed. S.~{D'Odorico}, 81--92

\bibitem[{{McCook} \& {Sion}(1999)}]{1999ApJS..121....1M}
{McCook}, G.~P., \& {Sion}, E.~M. 1999, \apjs, 121, 1

\bibitem[{{McLeod} {et~al.}(2015){McLeod}, {Geary}, {Conroy}, {Fabricant},
  {Ordway}, {Szentgyorgyi}, {Amato}, {Ashby}, {Caldwell}, {Curley}, {Gauron},
  {Holman}, {Norton}, {Pieri}, {Roll}, {Weaver}, {Zajac}, {Palunas}, \&
  {Osip}}]{2015PASP..127..366M}
{McLeod}, B., {Geary}, J., {Conroy}, M., {et~al.} 2015, \pasp, 127, 366

\bibitem[{{Miglio} {et~al.}(2012){Miglio}, {Brogaard}, {Stello}, {Chaplin},
  {D'Antona}, {Montalb{\'a}n}, {Basu}, {Bressan}, {Grundahl}, {Pinsonneault},
  {Serenelli}, {Elsworth}, {Hekker}, {Kallinger}, {Mosser}, {Ventura},
  {Bonanno}, {Noels}, {Silva Aguirre}, {Szabo}, {Li}, {McCauliff}, {Middour},
  \& {Kjeldsen}}]{2012MNRAS.419.2077M}
{Miglio}, A., {Brogaard}, K., {Stello}, D., {et~al.} 2012, \mnras, 419, 2077

\bibitem[{{Milone}(1992)}]{1992PASP..104.1268M}
{Milone}, A.~A.~E. 1992, \pasp, 104, 1268

\bibitem[{{Moehler} {et~al.}(2004){Moehler}, {Koester}, {Zoccali}, {Ferraro},
  {Heber}, {Napiwotzki}, \& {Renzini}}]{2004A&A...420..515M}
{Moehler}, S., {Koester}, D., {Zoccali}, M., {et~al.} 2004, \aap, 420, 515

\bibitem[{{Montgomery} {et~al.}(1993){Montgomery}, {Marschall}, \&
  {Janes}}]{1993AJ....106..181M}
{Montgomery}, K.~A., {Marschall}, L.~A., \& {Janes}, K.~A. 1993, \aj, 106, 181

\bibitem[{{Oke} {et~al.}(1995){Oke}, {Cohen}, {Carr}, {Cromer}, {Dingizian},
  {Harris}, {Labrecque}, {Lucinio}, {Schaal}, {Epps}, \&
  {Miller}}]{1995PASP..107..375O}
{Oke}, J.~B., {Cohen}, J.~G., {Carr}, M., {et~al.} 1995, \pasp, 107, 375

\bibitem[{{Origlia} {et~al.}(2006){Origlia}, {Valenti}, {Rich}, \&
  {Ferraro}}]{2006ApJ...646..499O}
{Origlia}, L., {Valenti}, E., {Rich}, R.~M., \& {Ferraro}, F.~R. 2006, \apj,
  646, 499

\bibitem[{{Osterbrock} {et~al.}(1996){Osterbrock}, {Fulbright}, {Martel},
  {Keane}, {Trager}, \& {Basri}}]{1996PASP..108..277O}
{Osterbrock}, D.~E., {Fulbright}, J.~P., {Martel}, A.~R., {et~al.} 1996, \pasp,
  108, 277

\bibitem[{{P{\^a}ris} {et~al.}(2014){P{\^a}ris}, {Petitjean}, {Aubourg},
  {Ross}, {Myers}, {Streblyanska}, {Bailey}, {Hall}, {Strauss}, {Anderson},
  {Bizyaev}, {Borde}, {Brinkmann}, {Bovy}, {Brandt}, {Brewington},
  {Brownstein}, {Cook}, {Ebelke}, {Fan}, {Filiz Ak}, {Finley}, {Font-Ribera},
  {Ge}, {Hamann}, {Ho}, {Jiang}, {Kinemuchi}, {Malanushenko}, {Malanushenko},
  {Marchante}, {McGreer}, {McMahon}, {Miralda-Escud{\'e}}, {Muna},
  {Noterdaeme}, {Oravetz}, {Palanque-Delabrouille}, {Pan}, {Perez-Fournon},
  {Pieri}, {Riffel}, {Schlegel}, {Schneider}, {Simmons}, {Viel}, {Weaver},
  {Wood-Vasey}, {Y{\`e}che}, \& {York}}]{2014AA...563A..54P}
{P{\^a}ris}, I., {Petitjean}, P., {Aubourg}, {\'E}., {et~al.} 2014, \aap, 563,
  A54

\bibitem[{{Pasquini} {et~al.}(1994){Pasquini}, {Belloni}, \&
  {Abbott}}]{1994AA...290L..17P}
{Pasquini}, L., {Belloni}, T., \& {Abbott}, T.~M.~C. 1994, \aap, 290

\bibitem[{{Perets} \& {Fabrycky}(2009)}]{2009ApJ...697.1048P}
{Perets}, H.~B., \& {Fabrycky}, D.~C. 2009, \apj, 697, 1048

\bibitem[{{Richer} {et~al.}(1998){Richer}, {Fahlman}, {Rosvick}, \&
  {Ibata}}]{1998ApJ...504L..91R}
{Richer}, H.~B., {Fahlman}, G.~G., {Rosvick}, J., \& {Ibata}, R. 1998, \apjl,
  504, L91

\bibitem[{{Rockosi} {et~al.}(2010){Rockosi}, {Stover}, {Kibrick}, {Lockwood},
  {Peck}, {Cowley}, {Bolte}, {Adkins}, {Alcott}, {Allen}, {Brown}, {Cabak},
  {Deich}, {Hilyard}, {Kassis}, {Lanclos}, {Lewis}, {Pfister}, {Phillips},
  {Robinson}, {Saylor}, {Thompson}, {Ward}, {Wei}, \&
  {Wright}}]{2010SPIE.7735E..0RR}
{Rockosi}, C., {Stover}, R., {Kibrick}, R., {et~al.} 2010, in \procspie, Vol.
  7735, Ground-based and Airborne Instrumentation for Astronomy III, 77350R

\bibitem[{Salaris {et~al.}(2009)Salaris, Serenelli, Weiss, \&
  Bertolami}]{Salaris2009}
Salaris, M., Serenelli, A., Weiss, A., \& Bertolami, M.~M. 2009, The
  Astrophysical Journal, 692, 1013

\bibitem[{{Sandquist} {et~al.}(2003){Sandquist}, {Latham}, {Shetrone}, \&
  {Milone}}]{2003AJ....125..810S}
{Sandquist}, E.~L., {Latham}, D.~W., {Shetrone}, M.~D., \& {Milone}, A.~A.~E.
  2003, \aj, 125, 810

\bibitem[{{Sarajedini} {et~al.}(2009){Sarajedini}, {Dotter}, \&
  {Kirkpatrick}}]{2009ApJ...698.1872S}
{Sarajedini}, A., {Dotter}, A., \& {Kirkpatrick}, A. 2009, \apj, 698, 1872

\bibitem[{Schlafly \& Finkbeiner(2011)}]{2011ApJ...737..103S}
Schlafly, E.~F., \& Finkbeiner, D.~P. 2011, \apj, 737, 103

\bibitem[{Slesnick {et~al.}(2002)Slesnick, Hillenbrand, \&
  Massey}]{Slesnick2002}
Slesnick, C.~L., Hillenbrand, L.~A., \& Massey, P. 2002, \apj, 576, 880

\bibitem[{{Steidel} {et~al.}(2004){Steidel}, {Shapley}, {Pettini},
  {Adelberger}, {Erb}, {Reddy}, \& {Hunt}}]{2004ApJ...604..534S}
{Steidel}, C.~C., {Shapley}, A.~E., {Pettini}, M., {et~al.} 2004, \apj, 604,
  534, reference for LRIS-B

\bibitem[{{Stello} {et~al.}(2016){Stello}, {Vanderburg}, {Casagrande},
  {Gilliland}, {Silva Aguirre}, {Sandquist}, {Leiner}, {Mathieu}, \&
  {Soderblom}}]{2016ApJ...832..133S}
{Stello}, D., {Vanderburg}, A., {Casagrande}, L., {et~al.} 2016, \apj, 832, 133

\bibitem[{{Stetson}(1987)}]{1987PASP...99..191S}
{Stetson}, P.~B. 1987, \pasp, 99, 191

\bibitem[{{Sweigart}(1994)}]{1994ApJ...426..612S}
{Sweigart}, A.~V. 1994, \apj, 426, 612

\bibitem[{{Tody}(1986)}]{1986SPIE..627..733T}
{Tody}, D. 1986, in \procspie, Vol. 627, Instrumentation in astronomy VI, ed.
  D.~L. {Crawford}, 733

\bibitem[{{Tody}(1993)}]{1993ASPC...52..173T}
{Tody}, D. 1993, in Astronomical Society of the Pacific Conference Series,
  Vol.~52, Astronomical Data Analysis Software and Systems II, ed. R.~J.
  {Hanisch}, R.~J.~V. {Brissenden}, \& J.~{Barnes}, 173

\bibitem[{{Tremblay} {et~al.}(2011){Tremblay}, {Bergeron}, \&
  {Gianninas}}]{2011ApJ...730..128T}
{Tremblay}, P.-E., {Bergeron}, P., \& {Gianninas}, A. 2011, \apj, 730, 128

\bibitem[{{Tremblay} {et~al.}(2013){Tremblay}, {Ludwig}, {Steffen}, \&
  {Freytag}}]{2013A&A...552A..13T}
{Tremblay}, P.-E., {Ludwig}, H.-G., {Steffen}, M., \& {Freytag}, B. 2013, \aap,
  552, A13

\bibitem[{{Tremblay} {et~al.}(2012){Tremblay}, {Schilbach}, {R{\"o}ser},
  {Jordan}, {Ludwig}, \& {Goldman}}]{2012A&A...547A..99T}
{Tremblay}, P.-E., {Schilbach}, E., {R{\"o}ser}, S., {et~al.} 2012, \aap, 547,
  A99

\bibitem[{{van den Berg} {et~al.}(2004){van den Berg}, {Tagliaferri},
  {Belloni}, \& {Verbunt}}]{2004AA...418..509V}
{van den Berg}, M., {Tagliaferri}, G., {Belloni}, T., \& {Verbunt}, F. 2004,
  \aap, 418, 509

\bibitem[{{van Dokkum}(2001)}]{2001PASP..113.1420V}
{van Dokkum}, P.~G. 2001, \pasp, 113, 1420

\bibitem[{{van Loon} {et~al.}(2008){van Loon}, {Boyer}, \&
  {McDonald}}]{2008ApJ...680L..49V}
{van Loon}, J.~T., {Boyer}, M.~L., \& {McDonald}, I. 2008, \apjl, 680, L49

\bibitem[{{VandenBerg} \& {Stetson}(2004)}]{2004PASP..116..997V}
{VandenBerg}, D.~A., \& {Stetson}, P.~B. 2004, \pasp, 116, 997

\bibitem[{{Williams}(2002)}]{Williams2002}
{Williams}, K.~A. 2002, PhD thesis, UNIVERSITY OF CALIFORNIA, SANTA CRUZ

\bibitem[{{Williams} {et~al.}(2009){Williams}, {Bolte}, \&
  {Koester}}]{2009ApJ...693..355W}
{Williams}, K.~A., {Bolte}, M., \& {Koester}, D. 2009, \apj, 693, 355

\bibitem[{Williams {et~al.}(2013)Williams, Howell, Liebert, Smith, Bellini,
  Rubin, \& Bolte}]{2013AJ....145..129W}
Williams, K.~A., Howell, S.~B., Liebert, J., {et~al.} 2013, \aj, 145, 129

\bibitem[{{Williams} {et~al.}(2006){Williams}, {Liebert}, {Bolte}, \&
  {Hanson}}]{2006ApJ...643L.127W}
{Williams}, K.~A., {Liebert}, J., {Bolte}, M., \& {Hanson}, R.~B. 2006, \apjl,
  643, L127

\bibitem[{Williams {et~al.}(2015)Williams, Serna-Grey, Chakraborty, Gianninas,
  \& Canton}]{2015AJ....150..194W}
Williams, K.~A., Serna-Grey, D., Chakraborty, S., Gianninas, A., \& Canton,
  P.~A. 2015, \aj, 150, 194

\end{thebibliography}

\end{document}